\documentclass[12pt]{report}
\usepackage{psfig, amsmath, amssymb}

\renewcommand{\baselinestretch}{1.5}

\begin{document}
\ifx\href\undefined\else\hypersetup{linktocpage=true}\fi

\def\cge      {{$_ >\atop{^\sim}$}}
\def\cle      {{$_ <\atop{^\sim}$}}
\def\etal     {{et\thinspace al. }}
\def\cg       {{c.\thinspace g.}}
\def\SN       {{$S/N$} }
\def\Ho       {{$H_{0}$} }
\def\qo       {{$q_{0}$} }
\def\Lstar    {{$L^{*}$} }
\def\magarc   {{\ mag\ arcsec$^{-2}$} }

\begin{titlepage}
\begin{center}
\setlength{\topmargin}{0cm}

\vspace{3cm}
{\bf {\LARGE A Determination of the Chemical Composition of {$\alpha$} 
Centauri A from Strong Lines}} \\
\vspace{3cm}
{\bf {\large Nimish Pradumna Hathi}} \\
{\large Department of Physics} \\
{\large The University of Queensland} \\
{\large Brisbane, Australia.} \\
\vspace{2cm}
{\bf {\large Supervisor: Dr. B. J. O'Mara}}\\
\vspace{2cm}
{\large Submitted: April 1997} \\
{\large Accepted: August 1997} \\
\vspace{2cm}
{\large A thesis submitted to the University of Queensland in \\
fulfilment of the requirements for the degree of Master of Science}

\end{center}
\end{titlepage}

\pagebreak

\tableofcontents

\setlength{\topmargin}{-0.25cm}
\chapter*{Abstract}
\addcontentsline{toc}{chapter}{Abstract}
The abundance of iron, magnesium, calcium and sodium in {$\alpha$} Centauri A ({$\alpha$} Cen
A) is determined from strong lines of these elements using the theory of collisional
line broadening of alkali and other spectral lines by neutral hydrogen developed by
Anstee \& O'Mara [1] and the spectrum synthesis program of John Ross [2]. The
derived abundance is independent of non--thermal motions in the photosphere and is
in good agreement with the results obtained by Chmielewski \etal [3] and Furenlid \&
Meylan [4] within the uncertainties involved.

As the abundances are derived from the wings of strong lines they are independent
of non--thermal motions in the photosphere of {$\alpha$} Cen A. The microturbulence is then
determined using medium--strong lines of iron by fixing the abundance of iron to the
value determined from the strong lines of iron, and adjusting the microturbulence
to match the observed equivalent width of each line. The microturbulence is the
average of the values obtained for all lines. This microturbulence can then be used
in determination of the abundance of other elements for which there are no strong 
lines in the spectrum.

The abundances of iron, magnesium, calcium and sodium in {$\alpha$} Cen A are approximately
0.2 dex higher than in the Sun. Also the microturbulence in the atmosphere of {$\alpha$}
Cen A which is (with the aid of the improved theory of line broadening) determined
independently of the abundances is 1.34 km/s i.e., 0.20 km/s greater than
on the Sun.

\chapter{Introduction}
A sound procedure for the determination of chemical composition of the Sun and stars
is important if we are to understand how elements are synthesised by thermonuclear
burning in stellar interiors, and how, in turn, this relates to the chemical evolution
of galaxies. The Sun and stars with similar surface temperatures are important as
lines of neutral and singly ionised metals are prevalent in these stars allowing abundances
to be determined for a wide selection of elements. The chemical composition
of the Sun is particularly important as it is used as a standard to which all other
stars are referenced. The lines best suited to abundance determinations are strong
lines as their wings are not affected by uncertainties in the line of sight motions
(microturbulence) of the absorbing atoms and their intrinsic strengths (f--values) can be
accurately determined in the laboratory.

In the past, a major impediment to the use of such lines has been the lack of an
adequate theory for the broadening of these lines by collisions with neutral hydrogen
atoms. Recently Anstee \& O'Mara (1991) [1] developed a theory which they tested
by applying it to the sodium D--lines in the solar spectrum. Anstee \& O'Mara (1995)
[5] used the theory to construct tables of line broadening cross--sections for s--p, p--s
transitions covering a range of effective principal quantum numbers for the upper
and lower levels of the transition. Anstee, O'Mara \& Ross (1997) [9] used these data
in the determination of the solar abundance of iron from the strong lines of Fe I.
The derived abundance was in excellent agreement with the meteoritic value ending
a long standing controversy about possible deviations of the solar abundance from
the meteoritic value. As a result of this work abundances can now be derived with
some confidence from strong lines with the resulting abundance being independent of
non--thermal motions in the photosphere.

In the past medium--strong lines have often been used whose equivalent widths
are sensitive to non--thermal motions in the stellar photosphere, and consequently it
has been difficult to determine whether the chemical composition of a star like $\alpha$
Cen A differs from that of the Sun. For example, Bessell (1981) [10] obtained a solar
composition for this star assuming a microturbulence of 1.7 km/s while more recently
Furenlid \& Meylan (1990) [4] and Chmielewski \etal (1992) [3] in an analysis of {$\alpha$}
Cen A \& B using a microturbulence of 1.0 km/s find that the abundance of iron
is systematically greater than in the Sun by 0.12 and 0.22 dex respectively. On the
other hand, if strong lines are used an abundance of iron can be derived which is independent 
of any assumed microturbulence. Moreover, once the abundance of iron has
been determined in this manner, the microturbulence can be uniquely determined by
requiring the abundance derived from medium--strong lines match that derived from
the strong lines. In this way the derived abundance and the microturbulence can be
decoupled leading to independent values for both. The derived microturbulence can
then be used in the determination of the abundance of elements for which there are
no strong lines in the star. It is the purpose of this thesis to test this new method of
analysis by application to the chemical composition of {$\alpha$} Cen A.

A brief review of the thesis contents:

In Chapter 2, the literature review of work on {$\alpha$} Cen is presented which
includes its physical properties and results obtained for its chemical composition.

In Chapter 3, the theory of the broadening of spectral lines by collisions with
atomic hydrogen is reviewed, with emphasis on the more recent work of Anstee \&
O'Mara (1991) [1] and Anstee \& O'Mara (1995) [5].

In Chapter 4, the spectrum synthesis software used in the thesis is applied to
selected medium--strong lines of Fe I in the solar spectrum to deduce the microturbulence
required to bring the abundance derived from strong and medium--strong lines
in the solar spectrum into agreement; a preview of the analysis to be carried out on
{$\alpha$} Cen A.

In Chapter 5, the {$\alpha$} Cen system is described and the spectra obtained for {$\alpha$} Cen A
\& B are described along with the process of reducing these data for analysis by the
spectrum synthesis program and identification of the lines.

In Chapter 6, abundances are derived from the strong and medium--strong lines in
{$\alpha$} Cen A and a microturbulence is derived.

In Chapter 7, conclusions of the analysis carried out in this thesis are summarized
and suggestions are made for future work.

\chapter{Literature Review}
The nearest star to our solar system, {$\alpha$} Cen, is in fact a system of three stars viz {$\alpha$}
Cen A, {$\alpha$} Cen B and Proxima Cen. The star system {$\alpha$} Cen, one of the pointers to
the Southern Cross, comprises of a --0.3 magnitude visual binary consisting of {$\alpha$} Cen
A, a G2V star (very similar to the Sun) and {$\alpha$} Cen B a K0V star. The orbital period
is 80.089 years, the semi--major axis is 23.5 AU, the distance 1.33 pc, the masses are
1.11 and 0.92 ($\pm$ 0.03)$M_\odot$. The third star Proxima Cen is closer to the Sun than
any other star. They are all about 4 light years away. They lie in the constellation
Centaurus, in the Milky Way and are visible from the Southern Hemisphere.

No less than 229 bibliographic references are given for {$\alpha$} Cen A in the SIMBAD
(Set of Identification, Measurements and Bibliography for Astronomical Data 1991
edition) and 140 references for {$\alpha$} Cen B [3]. According to SIMBAD the first detailed
analyses of {$\alpha$} Cen A and {$\alpha$} Cen B were done only in 1970 by French and Powell [11].

Recent detailed investigations of composition, temperature and ages of {$\alpha$} Cen A
and B have been in agreement in their conclusion that {$\alpha$} Cen A has temperature a
little higher or nearly equal to the Sun, is more metal--rich and is of the same age as
or older than the Sun (note that older stars usually have a lower metallicity than the
Sun). Exceptions are Bessell [10], whose analysis gives {$\alpha$} Cen A and B having the
same abundance as the Sun and Furenlid \& Meylan [4] concluded that the effective
temperature of {$\alpha$} Cen A is 5710 K, which is 70 K less than that of the Sun (here
adopted solar effective temperature is 5780 K).

Here is a brief summary of what some investigators have to say about {$\alpha$} Cen A.

Blackwell \& Shallis [12] describe the infrared flux method of determining stellar
angular diameters. The accuracy of the method is tested on Arcturus and twenty
seven other stars including {$\alpha$} Cen A. Given the angular diameter, $\theta$, the effective
temperature, $T_e$, is derived from the integrated flux from the star at the earth, $F_E$ ,
through the defining relation
\begin{equation}
\sigma T^4_e = 4*F_E/\theta^2,
\end{equation}
or alternatively given $F_E$ and the flux ($T_e$) from a suitable model atmosphere, $\theta$ can
be obtained. Their result for {$\alpha$} Cen A, using the adopted effective temperature of
5800 K, gives an angular diameter of $(86.2 \pm 2.3)*10^{-4}$ arcsec. The advantage of this
method is that it is applicable to cool stars, if a reasonably good model atmosphere
is available.

Bessell [10], using a grid of unpublished line blanketed models of Gustafsson, Bell,
Nordlund, Eriksson (1979) analysed Fe and Ti lines to obtain the abundance of these
elements in {$\alpha$} Cen A (and B). In his analysis Bessell [10] obtained abundance of
Fe and Ti in {$\alpha$} Cen A equal to the solar values. Two important parameters in this
analysis were found to have higher values. One was the adopted effective temperature
of 5820 K which is 40 K higher than the effective temperature of the Sun and the other
was a higher microturbulence of 1.7 km/s, which Bessell [10] concluded must have
contributed to above solar equivalent widths of lines in {$\alpha$} Cen A. Other investigators
do not agree with Bessell [10] regarding this higher microturbulence.

Soderblom and Dravins [13] measured the lithium abundance in {$\alpha$} Cen A. On
a scale where log N(H) = 12.00, they found a lithium abundance of log N(Li) =
1.28, which is nearly twice that of the Sun. As it is difficult to estimate the age
of a star from the lithium abundance alone, they assume that the age of {$\alpha$} Cen A,
depending on the composition, could be as low as 4 Gyr, or as great as 8 Gyr. In
this analysis Soderblom and Dravins [13] support an effective temperature of {$\alpha$} Cen
A which is same as the Sun. Their conclusion that {$\alpha$} Cen A has twice the lithium
abundance compared to the Sun is consistent with its greater mass (1.1$M_\odot$), despite
their prediction of probable evolutionary age of {$\alpha$} Cen A to be 6 Gyr.

Soderblom [14], later in 1986, concluded that there is a good agreement between
the (wings of) H$\alpha$ profiles of {$\alpha$} Cen A and the Sun which suggests they have 
the same effective temperature. Using this temperature and adopted luminosities he calculated
the radius of {$\alpha$} Cen A as 1.23$R_\odot$.

A differential abundance analysis between {$\alpha$} Cen A and the Sun has been carried
out by Furenlid \& Meylan [4]. They have measured around 500 lines of 26 elements
in very high \SN ratio (500--10,000) spectra for both objects. They find
that the absorption lines in {$\alpha$} Cen A are on average around 10 m{\AA} stronger than
the Sun. In their analysis Furenlid \& Meylan [4] found that there is an average
metal overabundance in {$\alpha$} Cen A of around 0.12 dex ($\pm$ 0.02 to $\pm$ 0.04 dex) on scale
of log N(H) = 12.0. The average abundance of Fe, Mg and Si relative to the Sun
was found to be 0.20 dex, using an effective temperature of 5800 K, log {\it g} = 4.0 and
microturbulence $\xi$ = 1.0 km/s. The effective temperature is reduced to 5710 K and the
overabundance of Fe reduces from 0.20 to 0.11 dex, when effects of the abundance on
continuous absorption are taken into account. It is unusual to find so low an effective
temperature as no other investigator has come up with this conclusion. Furenlid \&
Meylan [4] determined age of {$\alpha$} Cen A to be 4.2 $\pm$ 0.4 billion years which is nearly
same as that of the Sun.

A detailed abundance study for {$\alpha$} Cen A and $\alpha$ Cen B was carried out by Chmielewski
\etal [3]. Using high quality spectra with high \SN ratio and high resolution, 
they have estimated the effective temperature, iron abundance and lithium
abundance. The values they obtained are as follows : $T_{eff}$ = 5800 K, [Fe/H] = +0.22
$\pm$ 0.02 and log N(Li) = 1.4 $\pm$ 0.3. Their observations of the Ca II infrared triplet
indicate a low level of chromospheric activity for {$\alpha$} Cen A and B compared to the Sun.
From the abundance analysis they are able to confirm that {$\alpha$} Cen A is a Metal--Rich
(MR) star.

Chmielewski \etal [3] have summarised recent studies on {$\alpha$} Cen A. Table 2.1
summarises important analysis results for {$\alpha$} Cen A (here results of Chmielewski \etal
[3] are also included).

\renewcommand{\baselinestretch}{1}
\renewcommand{\arraystretch}{1.25}
\begin{table}[h]
\begin{center}
\caption{\it Results from Recent Analysis for {$\alpha$} Cen A.}
\begin{tabular}{|l|l|l|l|l|} \hline
Analysis & T{$_{eff}$} K & log {\it g} & $\xi$ km/s & [Fe/H] dex \\ \hline
$[11]$ & 5770 $\pm$ 70 & -- & -- & +0.22 $\pm$ 0.05 \\
$[10]$ & 5820 & 4.25 & 1.7 & --0.01 \\
$[13]$ & 5770 $\pm$ 20 & -- & -- & -- \\
$[4]$  & 5710 $\pm$ 25 & 4.0 $\pm$ 0.2 & 1.0 & +0.12 $\pm$ 0.02 \\
$[3]$  & 5800 $\pm$ 20 & 4.31 $\pm$ 0.02 & 1.0 & +0.22 $\pm$ 0.02 \\ \hline
\end{tabular}
\end{center}
\end{table}
\renewcommand{\baselinestretch}{1.5}

The metallicity of {$\alpha$} Cen A has been discussed by Noels \etal [15], Neuforge [16]
and Fernandes \& Neuforge [17]. All three papers have used different calibration
techniques and predicted that the metallicity is high in {$\alpha$} Cen A compared to the Sun
by a factor of 2 (0.3 dex). Noels \etal [15] using a generalisation of the solar calibration
technique computed evolutionary sequences for 1.085$M_\odot$ with different values of Z,
Y, and $\alpha$. They found that, the fraction by mass of elements more massive than He,
Z = 0.04, the helium mass fraction, Y $\sim$ 0.32 and convection parameter $\alpha$ = 1.6,
where $\alpha$ is the ratio of the mixing length to the pressure scale height i.e.$(l/H_p)$. The
helium abundance is higher than that of the Sun, which is nearly equal to 0.28. They
adopted mean values of effective temperature as 5765 K and Luminosity log (L/$L_\odot$) = 0.1853.

Neuforge [16] used the same calibration technique as that of Noels \etal [15] but
used different opacities and effective temperature. She computed her own set of low
temperature atmospheric opacities for different values of Z ranging from 0.02 to 0.04.
She adopted the effective temperature from Chmielewski \etal [3] which is 5800 K $\pm$
20 K. Neuforge [16] found the values for metallicity Z = 0.038, helium Y = 0.321,
age t = 4.84 Gyr and convection parameter $\alpha$ = 2.10. The value of the convection
parameter $\alpha$ is different from Noels \etal [15] only because of different opacities.

Fernandes and Neuforge [17] discuss two calibration techniques for fixed Z and
varied Z, using models calculated with mixing length convection theory (MLT) and
explain their solution through the behaviour of the convection parameter $\alpha$ with
chemical composition. They also predicted a higher Z for {$\alpha$} Cen A compared to the
Sun from their analysis.

\chapter{Theory of Spectral Line Broadening}
\section{Introduction to Line Broadening}
Broadening of spectral lines in stellar atmospheres is mainly due to three mechanisms.
They are:
\begin{itemize}
\item Natural broadening, due to the finite lifetimes of the upper and lower levels.
\item Collisional broadening, caused by perturbations of atomic energy levels during
collisions with other particles. There are two main sources of collisional
broadening in solar--type stellar atmospheres viz :
\begin{itemize}
\item Stark broadening, due to collisions with electrons and
\item so called van der Waals broadening, due to collisions with atoms, mainly
neutral hydrogen.
\end{itemize}
\item Doppler broadening, caused by random motions of the absorbing and emitting
atoms along the line of sight. These random motions may be due to:
\begin{itemize}
\item thermal motions of the atoms in the gas (thermal broadening) or
\item non--thermal random motions due to some form of small scale turbulence
(microturbulence).
\end{itemize}
\end{itemize}
Natural broadening and Doppler broadening are always present and dominate the
shapes near the line centre at low densities. However, natural broadening is important
only in the cores of strong lines formed in the upper layers of the stellar atmospheres.
The line shapes are strongly influenced by interactions of the radiating atoms or ions
with surrounding particles. Then the Collisional broadening (Pressure broadening)
comes into the picture. The most important interactions are those between radiating
systems and electrons. Because electric fields are involved, this type of broadening
is called Stark broadening. In the atmospheres of hot stars, electron collisions are
the most important because electrons are the dominant particle species, they move
rapidly and interact strongly with the absorbing atoms. On the other hand, the
outer layers of the Sun and stars of similar effective temperature are sufficiently hot
to dissociate molecular hydrogen into atomic hydrogen but are insufficiently hot to
ionize hydrogen. The few electrons present result from thermal ionisation of metals
such as iron and as a result of the low abundance of the metals, hydrogen atoms
outnumber electrons by about 10,000:1. The collisional broadening of absorption
lines in these stars is therefore dominated by collisions with atomic hydrogen.

In addition to the above classification of the various pressure broadening mechanisms, 
there is the division into impact (collision) and quasi--static (statistical) broadening. 
These are two extreme approximations in the general theory of broadening.
The quasi--static description is valid when the perturbers move relatively slowly, so
that the perturbation is practically constant over the time of interest, which is, at
most, of the order of the inverse of the line width, the latter being measured in angular
frequency units. In the impact approximation, on the same time scale, the duration
of individual collisions is negligibly small which is always the case for broadening by
collisions with electrons and hydrogen atoms.

As collisions with hydrogen atoms dominate the collisional broadening of lines
in cool stars like the Sun, it is important to have a satisfactory theory for this line
broadening process if strong lines are to be used in determination of the chemical
composition of these stars.

\section{Broadening by Collisions with Hydrogen Atoms}
A theory for the broadening of spectral lines by collisions with atomic hydrogen developed 
by Anstee \& O'Mara [1] is described in a paper presented by O'Mara [18] at
an IAU Conference in Sydney in January 1997. An extract from that paper is quoted
below. 
\begin{quote}
``In cool stars neutral hydrogen atoms outnumber electrons by four
orders of magnitude, consequently the broadening of most spectral lines
is dominated by collisions with neutral hydrogen atoms. Conventional vander 
Waals' theory for this broadening process is known to underestimate
the broadening of spectral lines in the Sun by about a factor of two.
Development of a satisfactory theory is important as it would allow strong
lines with well determined f--values to be used to determine abundances
in cool stars in a manner which is independent of photospheric motions.
Also such lines could also be used to determine surface gravities in cool
stars.

Collisions with neutral hydrogen atoms are sufficiently fast for the
impact approximation of spectral line broadening theory to be valid. In
this approximation the line has a Lorentz profile with a half half--width
which is given by:
\begin{equation}
w = N \int_0^\infty v f(v) \sigma(v) dv,
\end{equation}
where $N$ is the hydrogen atom number density, {\it v} is the relative collision
speed, {\it f(v)} is the speed distribution and the line broadening cross--section
\begin{equation}
\sigma(v) = 2\pi \int_0^\infty <\Pi(b,v)>_{av} bdb,
\end{equation}
where the integrand contains the product of the geometrical cross--section
2$\pi bdb$ and a line broadening efficiency factor $<\Pi(b,v)>_{av}$ for collisions
with impact parameter {\it b} and relative speed {\it v}.

$< ... >_{av}$ indicates that the efficiency factor has to be averaged over all
orientations of the perturbed atom. The efficiency factor can be expressed
in terms of the S--matrix elements for the collision which are functionally
dependent on the interaction energy between a hydrogen atom in the
ground state and the perturbed atom in its upper and lower states. The
only essential difference between various theoretical treatments is in the
method employed to determine this interaction energy.

In the theory developed by Anstee \& O'Mara [1] the interaction energy
is calculated using Rayleigh--Schr\"{o}dinger perturbation theory. If exchange
effects are neglected, the shift in energy of the two--atom system as a result
of the electrostatic interaction V between them is given by
\begin{equation}
\Delta E_i = <i|V|i> + \sum_{j \neq i} \frac{<i|V|j><j|V|i>}{E_i - E_j},
\end{equation}
where the unperturbed eigenstates of the two--atom system $|i>$ are products
of the unperturbed eigenstates of the two atoms. As first pointed out
by Uns\"{o}ld the above expression can be greatly simplified if $E_i - E_j$ can
be replaced by a constant value $E_p$ . Closure can then be used to complete
the sum over j to obtain
\begin{figure}[ht]
\hspace{2cm}
\psfig{file=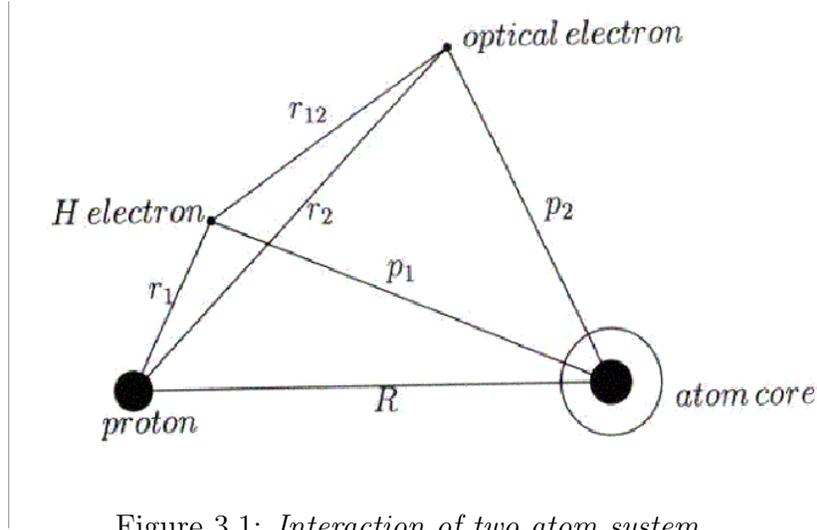,height=7.0cm,width=11.0cm}
\vspace{-1cm}
\caption{\it Interaction of two atom system.}
\end{figure}
\begin{equation}
\Delta E_i = <i|V|i> + \frac{1}{E_p} (<i|V^2|i>-<i|V|i>^2).
\end{equation}
The second term accounts for the interatomic interaction resulting from
fluctuations of both atoms simultaneously (dispersion) and the fluctuation
of each atom in the static field of the other (induction). The second term
dominates the interaction. To develop the theory further, the perturbed
atom is described by an optical electron outside a positively charged core
so that product states of the two--atom system have the form $|100>
|n^*lm>$. With reference to the Figure 3.1, the electrostatic interaction
energy V , in atomic units, is
\begin{equation}
V = \frac{1}{R} + \frac{1}{r_{12}} - \frac{1}{r_2} - \frac{1}{p_1}.
\end{equation}
For the state $|i>=|100>|n^*lm>$, the interaction energy can be
expressed in the form
\begin{eqnarray}
\Delta E_{n^*l|m|} = <i|V|i> + \frac{1}{E_p} \int_0^\infty R^2_{n^*l}(p_2) 
I_{l|m|}(p_2,R) p^2_2 dp_2 \\  
- \frac{1}{E_p} <i|V|i>^2, \nonumber
\end{eqnarray}
where $R_{n^*l}(p_2)$ is the radial wave function for the optical electron in the
perturbed atom and $I_{l|m|}(p_2,R)$ are lengthy complicated analytic functions
of $p_2$ and R, which have a logarithmic singularity at $p_2$ = R and which
can be expressed as an asymptotic expansion in powers of $\frac{1}{R^2}$
when R is large. It can be shown that the leading term in this expansion leads to
\begin{equation}
<\Delta E_{nl|m|}>_{av} \sim \frac{1}{E_p} \frac{<p^2_2>}{R^6},
\end{equation}
and if $E_p$ is chosen to be 4/9 atomic units 1/$E_p$ = 9/4 which is the
polarizability of hydrogen in atomic units. Thus
\begin{equation}
<\Delta E_{nl|m|}>_{av} \sim \alpha_H \frac{<p^2_2>}{R^6},
\end{equation}
the standard expression for the van der Waals interaction between the two
atoms. However, the impact parameters important in the determination
of the cross--section are always too small for this asymptotic form of the
interaction to be valid. The terms in $<i|V|i>$ can be expressed in a
similar but simpler form. It is an important feature of the method that the
interaction energy between the two atoms can be determined analytically
to within a numerical integration over the radial wave function for the
perturbed atom.

For individual transitions of interest Scaled Thomas--Fermi--Dirac or
Hartree--Fock radial wave functions can be used in the determination of
the interaction energy. Standard methods can then be used to determine
the efficiency factor $<\Pi(b,v)>_{av}$ and these can be used to calculate
the cross--section, and ultimately, the line width. This is perhaps the
best method for specific lines of interest such as the Na D--lines and the
Mg b--lines. However, without significant loss of accuracy, Coulomb wave
functions can be used to tabulate cross--sections for a range of effective
principal and azimuthal quantum numbers for the upper and lower levels 
of the transition. This approach enables cross--sections to be obtained
for a wide variety of transitions by interpolation. Anstee \& O'Mara [5]
adopted this approach for s--p and p--s transitions. In addition to tabulating
cross--sections for a collision speed of $v_0 = 10^4 m/s$ for a range of
effective principal quantum numbers for the upper and lower level, they
also determined by direct computation, velocity exponents $\alpha$, on the 
assumption that $\sigma(v)\sim v^{-\alpha}$.

With this dependence of the cross--section on collision speed the
integration over the speed distribution can be performed to obtain the line
width per unit H--atom density, which is given by 
\begin{equation}
\frac{w}{N} = \left(\frac{4}{\pi}\right)^{\alpha/2} \Gamma \left(\frac{4-\alpha}{2}\right) 
v_0 \sigma(v_0) (\bar{v}/v_0)^{1-\alpha},
\end{equation}
where $\bar{v}=\left(\frac{8kT}{\pi \mu}\right)^{1/2}$, and $\mu$ is the reduced mass of 
the two atoms. Typically
$\alpha$ is about 0.25, which leads to a temperature dependence of $T^{0.38}$ for the
line width. At present, tabulated values of $\sigma$ and $\alpha$ are only available for
s--p and p--s transitions but work is in progress to extend the results to
p--d, d--p and d--f, f--d transitions.''
\end{quote}
The relation between this treatment of broadening by collisions with hydrogen atoms
and others is discussed by Anstee \& O'Mara [1] and Anstee [19].

\section{Conclusions}
It is now no longer necessary to use the time honoured van der Waals' theory to
calculate broadening cross--sections for collisions with hydrogen atoms which are known
to underestimate the broadening by a factor of two. The tabulated data of Anstee
\& O'Mara [5] can be used to obtain cross--sections for s--p, p--s transitions. New data
for p--d, d--p transitions have been submitted for publication by Barklem \& O'Mara
[20] and work is in progress for d--f, f--d transitions. Anstee \& O'Mara [5] have shown
that these data yield solar abundances from selected strong lines which are consistent
with meteoritic values and Anstee, O'Mara \& Ross [9] have shown that strong lines
can now be used to obtain the solar abundance of iron with some precision. The iron
abundance obtained is in detailed agreement with the meteoritic value, ending a long
standing controversy concerning the solar abundance of iron.

The only remaining significant approximation in this theoretical work is the Uns\"{o}ld
approximation. In future it is hoped to improve on this approximation by forcing the
potentials to match more accurate potentials obtained by other means at the long
range van der Waals limit.

These data are used in this thesis to obtain abundances, which are independent
of any assumed microturbulence, in {$\alpha$} Cen A from strong lines with well determined
f--values.

\chapter{The Determination of the Microturbulence from
Medium--Strong Lines of Fe I in the Solar Spectrum}
\section{Introduction}
Whether {$\alpha$} Cen A is predicted to be metal rich or not is critically dependent on
the assumed microturbulent velocity $\xi$ in the photosphere of this star. For example
Bessell [10] finds that {$\alpha$} Cen A has the same composition as the Sun if $\xi$ = 1.7
km/s while Furenlid \& Meylan [4] find it to be metal rich if $\xi$ = 1.0 km/s. The
improved line broadening theory of Anstee \& O'Mara [1] permits the determination
of the composition from the wings of strong lines in a manner which is independent
of the microturbulence. The microturbulence can then be independently determined
by requiring that the composition derived from medium-­strong lines match the abundance 
derived from the strong lines. We illustrate this procedure by application to
the solar spectrum as a prelude to its application to {$\alpha$} Cen A.

\section{Microturbulence Analysis of Oxford Fe I Lines}
Holweger [21] discusses meteoritic and solar abundances 
and their stability as cosmic
reference data. During his discussion, Holweger [21] plots log N(Fe) versus Equivalent 
Width (m{\AA}) using a standard damping enhancement factor of 2.0 and zero
microturbulence. This plot is shown in Figure 4.1.

\begin{figure}[h]
\hspace{2cm}
\psfig{file=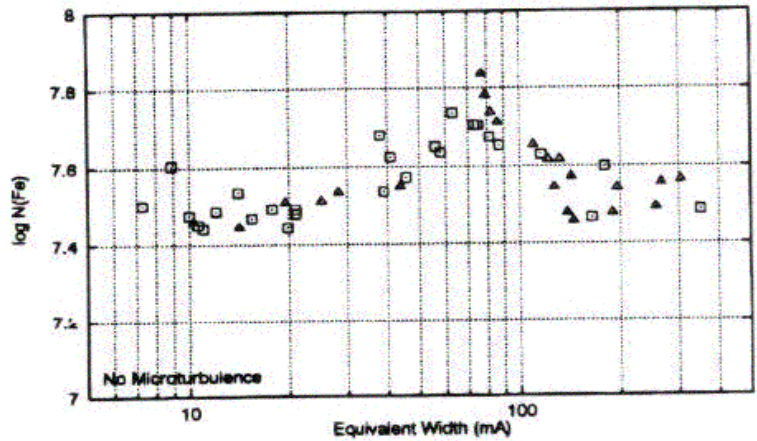,height=6.0cm,width=10.0cm}
\caption{\it log N(Fe) versus Equivalent widths using zero microturbulence.}
\end{figure}

Omitting microturbulence altogether introduces a conspicuous hump in the abundance 
pattern; most sensitive are lines of intermediate strength with equivalent widths
in the range 60-­100 m{\AA}. Figure 4.1 gives a clear picture of what happens if the presence 
of small scale motions in the stellar atmosphere is neglected in line formation
calculations.

In this section we have used the Fe I line data from Blackwell, Lynas-­Gray and
Smith [22] to study the effect of microturbulence, while keeping the abundance fixed
at the meteoritic value of 7.51. The results are shown in Table 4.1.

\renewcommand{\baselinestretch}{1}
\renewcommand{\arraystretch}{1.25}
\begin{table}[ht]
\begin{center}
\parbox{15cm}{\caption {\it Microturbulence Analysis of Oxford Fe I Lines. The line broadening 
cross­-sections $\sigma$ and velocity parameters {$\alpha$} were obtained by interpolation in the tables of
Anstee and O'Mara [5]. The microturbulence required to match the observed equivalent
widths were obtained using the spectrum synthesis program of Ross [2]. '*' indicates
lines which are weaker (less than 60 m{\AA}) and as weaker lines are not affected by
microturbulence we get very high values of microturbulence if we keep the abundance
fixed at 7.51.}}
\begin{tabular}{|l|l|l|l|l|l|} \hline
Wavelength & Equivalent & Broadening & Enhance- & Velocity & Microturbulence \\*[-0.15cm]
({\AA}) & Width & Cross- & ment & Parameter & ($\xi$ km/s) {\bf with} \\*[-0.15cm]
        & (m{\AA}) & section $\sigma$ & Factor & {$\alpha$} & {\bf an abundance} \\*[-0.15cm]
        &          & (a$^{2}_{0}$) &  &  & {\bf 7.51} \\ \hline
4389.24 & 71.70 & 218 & 2.13 & 0.250 & 1.010 \\
4445.47 & 38.80 & 219 & 2.15 & 0.250 & 2.300* \\
5247.06 & 65.80 & 206 & 3.30 & 0.254 & 1.220 \\
5250.21 & 64.90 & 208 & 3.22 & 0.254 & 1.240 \\
5701.55 & 85.10 & 364 & 2.05 & 0.241 & 1.110 \\
5956.70 & 50.80 & 229 & 2.54 & 0.252 & 1.695 \\
6082.71 & 34.00 & 240 & 1.99 & 0.249 & 2.170* \\
6137.00 & 63.80 & 282 & 1.62 & 0.266 & 0.952 \\
6151.62 & 48.20 & 280 & 1.62 & 0.264 & 1.148 \\
6173.34 & 67.40 & 283 & 1.62 & 0.267 & 1.025 \\
6200.32 & 75.60 & 354 & 2.20 & 0.240 & 1.148 \\
6219.29 & 91.50 & 280 & 1.63 & 0.264 & 0.940 \\
6265.14 & 86.80 & 277 & 1.63 & 0.262 & 1.065 \\
6280.63 & 62.40 & 225 & 2.85 & 0.253 & 1.305 \\
6297.80 & 75.30 & 280 & 1.63 & 0.264 & 1.012 \\
6322.69 & 79.20 & 348 & 2.25 & 0.242 & 1.316 \\
6481.88 & 64.20 & 312 & 2.47 & 0.246 & 1.311 \\
6498.95 & 44.30 & 228 & 2.93 & 0.254 & 2.500* \\
6574.24 & 26.50 & 229 & 2.96 & 0.254 & 9.000* \\
6593.88 & 86.40 & 324 & 2.43 & 0.248 & 1.154 \\
6609.12 & 65.50 & 337 & 2.37 & 0.246 & 1.267 \\
6625.04 & 13.60 & 230 & 2.98 & 0.254 & 13.00* \\
6750.15 & 75.80 & 282 & NA & 0.260 & 1.132 \\
6945.21 & 83.80 & 279 & NA & 0.261 & 1.076 \\
6978.86 & 80.10 & 283 & NA & 0.261 & 1.102 \\
7723.20 & 38.50 & 255 & NA & 0.260 & 5.300* \\ \hline
\end{tabular}
\end{center}
\end{table}
\renewcommand{\baselinestretch}{1.5}

As stated by Holweger [21] and described above, the lines most affected by 
microturbulence are lines of intermediate strengths between 60-­100 m{\AA}, hence we have
omitted weaker lines in our analysis. The variation of microturbulence for equivalent
widths in range of 60-­100 m{\AA} is between 0.9 to 1.3 km/s. The mean value is 1.14 $\pm$
0.12 km/s.

\section{Conclusions}
Using the lines most sensitive to microturbulence with equivalent widths in the range
of 60-­100 m{\AA} it has been shown that a microturbulence of 1.14 $\pm$ 0.12 km/s is
required if these lines are to yield the same abundance of iron as that obtained by
Anstee, O'Mara \& Ross [9] and in meteorites. This value is somewhat larger than
1.0 km/s, a value commonly employed in analysis of the solar spectrum. As {$\alpha$} Cen
A is very similar to the Sun, and as a similar method will be used to determine the
microturbulence in {$\alpha$} Cen A, this new value 1.14 km/s will be a useful benchmark for
comparison.

\chapter{Data Analysis for {$\alpha$} Cen A}
\section{{$\alpha$} Cen -- the Closest Stellar System}
{$\alpha$} Cen is of special interest as it is the closest stellar system to the Sun. $\alpha$
Cen lies 4.35 light-­years from the Sun. It is not a single star but is actually a triple
star system. Its brightest and warmest star is called {$\alpha$} Cen A. It is a yellow star with
spectral type of G2, exactly the same as that of the Sun. Therefore its temperature
($\sim$ 5800 K) and color are similar to those of the Sun. {$\alpha$} Cen A, the brightest
star in the system, is slightly more massive and luminous, its mass is 1.09 solar masses
and its brightness is 54 percent greater than that of the Sun.

{$\alpha$} Cen B, the second brightest star in the system, lies close to {$\alpha$} Cen A. It
is an orange star, cooler and smaller than the Sun. Its spectral type is K1 and its
surface temperature is 5300 K, approximately 500 K lower than that of the Sun. The
mass of {$\alpha$} Cen B is 0.90 solar masses and this star's brightness is just 44 percent of
the solar value. The brightest components {$\alpha$} Cen A and B form a binary. They orbit
each other in about 80 years with a mean separation of 23 astronomical units. Figure
5.1 shows the comparative size of {$\alpha$} Cen stars with the Sun, taken from Croswell [6].

The third and the faintest member of the {$\alpha$} Cen system, {$\alpha$} Cen C, lies a long way
\begin{figure}[ht]
\hspace{0.5cm}
\psfig{file=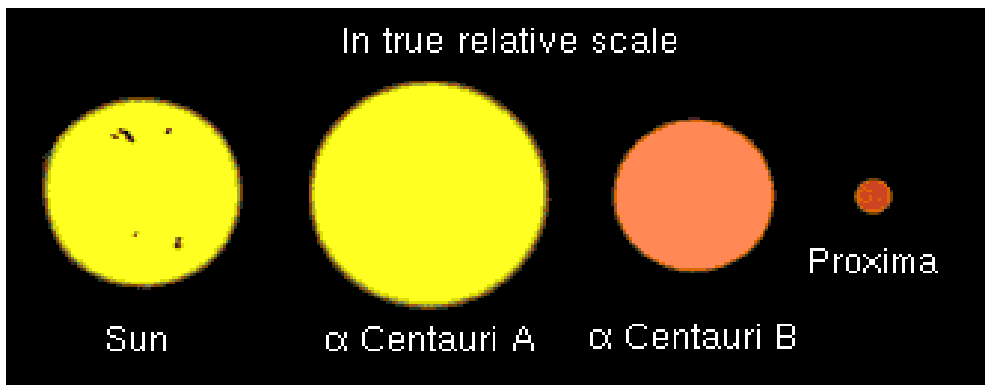,height=5.0cm,width=13.0cm}
\caption{\it Relative Size of {$\alpha$} Cen Stars compared to the Sun.}
\end{figure}
from its two brighter companions. {$\alpha$} Cen C lies 13,000 AU from A and B, or 400
times the distance between the Sun and Neptune. It is not yet known whether {$\alpha$} Cen
C is really bound to A and B, or whether it has left the system some million of years
ago. {$\alpha$} Cen C lies measurably closer to the Sun than the other two: it is only
4.22 light-­years away, and it is the nearest individual star to the Sun. Because of its
proximity, {$\alpha$} Cen C is also called Proxima Cen. Proxima Cen is a dim red dwarf,
much fainter, cooler and smaller than the Sun. Its spectral type is M5, temperature
is half that of the Sun, mass one-­tenth that of the Sun, and brightness a mere 0.006
percent that of the Sun [6]. Proxima is so faint that astronomers did not discover it
until 1915.

\section{Observation and Data}
The analysis of {$\alpha$} Cen A reported here is differential to the Sun and based on an
entirely new set of observational data. The observational material for {$\alpha$} Cen A consists
of high resolution, high \SN ratio spectra obtained at Mt. Stromlo and
Siding Spring Observatories (MSSSO), and the solar data are from the Kitt Peak
solar flux atlas of Kurucz \etal [23].

\renewcommand{\baselinestretch}{1}
\renewcommand{\arraystretch}{1.25}
\begin{table}[ht]
\begin{center}
\caption{\it Comparison of the Sun with {$\alpha$} Cen Stars [6].}
\begin{tabular}{|l|l|l|l|l|} \hline
The Sun and its & Sun & {$\alpha$} Cen A & {$\alpha$} Cen B & Proxima \\*[-0.15cm]
Nearest         &     &                  &                  &         \\*[-0.15cm]
Neighbours      &     &                  &                  &         \\ \hline
Color & Yellow & Yellow  & Orange & Red \\
Spectral Type & G2 & G2 & K1 & M5 \\
Temperature & 5770 K & 5800 K & 5300 K & 2700 K \\
Mass & 1.00 & 1.09 & 0.90 & 0.1 \\
Radius & 1.00 & 1.2 & 0.8 & 0.2 \\
Brightness & 1.00 & 1.54 & 0.44 & 0.00006 \\
Distance (light years) & 0.00 & 4.35 & 4.35 & 4.22 \\
Age (billion years) & 4.6 & 5-­6 & 5-­6 & 1? \\ \hline
\end{tabular}
\end{center}
\end{table}
\renewcommand{\baselinestretch}{1.5}
Spectral data for {$\alpha$} Cen A were taken by Dr. Mike Bessell at MSSSO. The spectra were 
taken on the 74 inch telescope at Mt. Stromlo Observatory in June and July 1996
with the echell\'{e} system. They were taken with the 120 inch focal length coud\'{e} camera
and a 31.6 groove/mm echell\'{e} cross dispersed with a 150 lines/mm grating. The CCD
was a SITe 2048x2048 CCD with 24 micron pixels. A fast rotating B star was also
observed for the red settings so that the atmospheric absorption lines due to water
and $O_2$ could be identified and removed. The \SN varied with order across the CCD
because of the vignetting due to the cross-­disperser not being near a pupil. The actual
resolution was about 3 pixels (these 1024 ''pixels'' were double binned pixels of the
2048 CCD pixels) leading to an effective resolving power of about 200,000. The CCD
had a gain of 2 electrons per ADU so most of the exposures were aimed at about
60000 electrons maximum. As the data were between 60000 and 180000 electrons per
resolution element, the nominal \SN ratio was between 200 to 400.

Around 100 spectral frames of {$\alpha$} Cen A with very high resolution were obtained,
giving almost continuous coverage from 4000--7000 {\AA}. These data were used for the
{$\alpha$} Cen A analysis.

The data were sent in two forms. One was formatted FITS files and second was in
a hard copy plot of the data. Unfortunately, the FITS files became unusable because
of errors in the transmission and hence ASCII files were used directly for data analysis.
The spectral frames were in blocks of nearly 16-­20 {\AA} and strong lines were used to
identify their wavelengths. A detailed analysis (formatting) of data was carried out
to make data usable for the spectral line synthesis program (SYN) developed by John
Ross [2]. The following sections describe how the procedure was carried out and what
modifications were required in the programs.

\section{Spectrum Synthesis Program (SYN)}
The Spectrum Synthesis program (SYN) developed by John Ross [2], is an adaptation
of a program first used by Ross \& Aller (1976) [24]. The synthesis program, which runs
on an IBM-­compatible PC with an 80486 DX processor and VGA graphics, is highly
interactive. The observed and computed spectra with the residual error multiplied
by a factor of 5 are displayed simultaneously. Parameters such as the collisional
damping constant, the abundance, the line wavelength, continuum level and so on,
can be changed at will to minimize the residual error. In addition, blending lines can
be added to the spectra in cases where they affect the fitting process. The software
is now accessible over the Internet [2].

\section{Identification of Spectral Lines}
Identification of lines was carried out using the solar spectrum identification list of
Moore \etal [7] because {$\alpha$} Cen A is similar to the Sun in composition as well as effective
temperature. Using the line identification list, the hard copy of the spectrum of {$\alpha$} Cen
A, and the solar flux atlas displayed on a computer it was easy to identify individual
lines in {$\alpha$} Cen A.

{$\alpha$} Cen A spectra were in spectral frames of approximately 20 {\AA} and it was
not continuous. The spectral data was basically in the range of 3700 {\AA} to 6000 {\AA}
and one additional separate range of 7000 {\AA} to 8000 {\AA} was also available. Identifying
major lines of Fe (iron), Mg (magnesium), Ca (calcium), etc., was done and a table
similar to the solar atlas was made ({$\alpha$} Cen atlas).

\section{Data Formatting for SYN}
The ASCII data files contained intensity values corresponding to channel numbers. 
Each file had 1024 data points and had a wavelength range of about 20 {\AA}.
A cubic least squares fitting process was used to obtain the relationship between
channel numbers and the wavelengths of identified lines (a process popularly known as
{\it scrunching the spectrum}). The values obtained were used in a FORTRAN program
to generate data files having intensity values corresponding to the wavelengths. The
data files with intensity and wavelengths had irregular wavelength intervals. The
data files were interpolated using MATLAB to a uniform wavelength interval of 0.01 {\AA}. 
Each data set was then scaled so that the highest point in the continuum was
9999, the value used in the spectrum synthesis program. Data files were merged to
form 100 {\AA} blocks, so, the file contained data from wavelength 3700.00 {\AA} to 3799.99 {\AA}, 
3800.00 {\AA} to 3899.99 {\AA} and so on. Finally these data files were converted from
ASCII to binary, the format required by the synthesis program SYN.

The {$\alpha$} Cen A spectrum was compared with the solar flux spectrum in the ATLAS
program (a program originally developed by J. E. Ross to compare different atlases
of the solar spectrum). During this comparison some spectral frames of {$\alpha$} Cen A
showed a 'tilt' towards either of the blue or red side of the spectrum. The ATLAS
program was modified to allow these tilts to be removed. The ATLAS program
also provides the facility to shift the spectrum both horizontally (wavelength) and
vertically (intensity) to help in straightening the spectra.

The above mentioned process and programming were carried out to satisfy the
requirements of the SYN program. The SYN program uses these binary files to plot
the observed spectrum. It also uses data files which contain parameters from the
chosen atmospheric model and atomic data for the lines to compute the theoretical
spectrum. The program then compares the observed and the theoretical
spectra to deduce the abundance of a particular element.

SYN requires an accurate atmospheric model for the comparison of the theoretical
spectrum and the observed spectrum. SYN contains a program for computing a
model atmosphere using as input data, the assumed temperature distribution, surface
gravity and composition. As the effective temperature of {$\alpha$} Cen A is close to that
of the Sun, the temperature distribution can be obtained by scaling the temperature
distribution of a solar model to the effective temperature distribution of {$\alpha$} Cen A.
The Holweger-­M\"{u}ller model [25] for the Sun was used for this purpose.

\chapter{Results and Discussion}
\section{Importance of Strong Lines in Determining Chemical Composition}
Weak, medium-­strong, and very strong lines can be used to determine the chemical
composition of a star. Weak lines have the advantage that their equivalent widths
are independent of all forms of broadening. However, because they are faint it is
difficult to measure their intrinsic strengths in the laboratory with any precision and
to observe them with sufficient \SN ratio in all stars other than the Sun. Also they
are very susceptible to the presence of weak unknown blends.

Medium-­strong lines having equivalent widths of about 80 m{\AA} present their own
problems. As pointed out by Holweger [21], the abundance derived from such lines is
particularly sensitive to non-­thermal motions (microturbulence) in the stellar photosphere. 
This sensitivity is largely responsible for the disparity between determinations
of the abundance of iron in the Sun by different groups.

Finally, the method which is used here and the method which is capable of estimating 
abundances to quite a degree of accuracy, is the use of very strong lines with
well developed damping wings for abundance determination. The problem which
weak lines face is solved here, as the strengths of these strong lines in laboratory
sources makes it easier to determine their f-­values with some precision. The problem
of non-­thermal motions is solved by the process of fitting the wings of these lines,
which excludes the line core, which is affected by non-­thermal motions of the atoms
and is more sensitive to the deviations from the local thermodynamic equilibrium
(LTE).

The major problem of using these strong lines for abundance determination in the
past was the lack of suitable empirical or theoretical damping constants. Now, with
the use of the tables of line-­broadening cross-­sections produced by Anstee \& O'Mara
[5] for s-­p and p-­s transitions, it is possible to obtain damping constants. Anstee \&
O'Mara [5] and Anstee, O'Mara \& Ross [9] have shown that these damping constants
with good f-­values for a selection of strong lines produce solar abundances that are
consistent with meteoritic abundances.

This method and these damping constants are used here to determine the abundance 
of selected elements in {$\alpha$} Cen A, a solar-­type star.

\section{Parameters for an Atmospheric Model}
There are four parameters of importance in determining the atmospheric model for
a star. They are effective temperature ($T_e$), surface gravity (log {\it g}), chemical 
composition and microturbulence ($\xi$). All four parameters, and how they are used in this
analysis, are described in brief below.

\subsection{Effective Temperature ($T_e$)}
The effective temperature ($T_e$) of {$\alpha$} Cen A is considered to be nearly same as
that of the Sun. Its color suggests a range from the solar effective temperature of
5780 K to 5820 K. In this analysis we have used two models with different effective
temperatures, one with the solar effective temperature 5780 K and the other, 5820 K,
obtained by Bessell [10]. The resultant abundance of the elements obtained in these
two cases should bracket the range for {$\alpha$} Cen A.

\subsection{Surface Gravity (log {\it g})}
The surface gravity (log {\it g}) is one of the most important parameters in estimating the
chemical composition of {$\alpha$} Cen A. Its variation has lead to differences in the exact
value of the overabundance in {$\alpha$} Cen A. Chmielewski \etal [3] found log {\it g} to be 4.31
$\pm$ 0.02 while Furenlid \& Meylan [4] found a low value for log {\it g} of 4.0 $\pm$ 0.2. To check
the value of log {\it g}, the angular diameter $86.2*10^{-4}$ arcsec from Blackwell \& Shallis
[12], the parallax of 0.7506'' from Furenlid \& Meylan [4], leads to
\begin{equation}
R_{\alpha CenA} = 1.234 R_\odot.
\end{equation}
This value of the radius is in agreement with the value of 1.23$R_\odot$ obtained by
Soderblom [14]. If the parallax is 0.754'', as obtained from Hipparcos data [26], the
radius comes out to be 1.228$R_\odot$ which is in agreement with both the above values
to within the uncertainties involved.

Using the above value of the radius, the mass of 1.085$M_\odot$ from Furenlid
\& Meylan [4] and log $g_\odot$ = 4.44 in the expression
\begin{equation}
g/g_\odot = (M/M_\odot) * 1/(R/R_\odot)^2
\end{equation}
we get,
\begin{equation}
log(g/g_\odot) = -0.14
\end{equation}
which gives,
\begin{equation}
log(g_{\alpha CenA}) = 4.30 \pm 0.04
\end{equation}
The estimated uncertainty of $\pm$0.04 is based on the uncertainty in the parallax
and in the radius. The lower surface gravity means the number density of hydrogen
atoms is less than in the Sun and hence the line broadening is reduced. This is shown
to be important later where inspite of the profiles of strong lines in the spectrum
of {$\alpha$} Cen A being nearly the same as the Sun, a higher abundance of the element is
required to make up for the reduced line broadening.

This value of surface gravity is kept fixed throughout the analysis.

\subsection{Microturbulence ($\xi$)}
Microturbulence ($\xi$) is also a very important parameter in the determination of abundances 
of elements in {$\alpha$} Cen A. It has been stated previously, in Chapter 4, that the
microturbulence is very important for lines having intermediate strengths of 60-­100
m{\AA}. In the first part of this analysis, strong lines of Fe, Mg, Ca and Na are used to
obtain the abundance of these elements in {$\alpha$} Cen A, the strong lines are not affected
by microturbulence. In the second part, medium-­strong lines of iron are used to determine 
microturbulence, by fixing the abundance of iron to the value obtained from
the first part, thus leading to an independent determination of the microturbulence
which can then be used in the determination of the abundance of other elements with
no strong lines in the star.

\subsection{Chemical Composition}
The fourth parameter, the chemical composition, is determined from the synthesis of
the strong lines. However, as the composition has an influence on the atmospheric
model several iterations may be necessary before the appropriate composition and
model are obtained.

\section{Chemical Composition Using Strong Lines}

\subsection{Selection of Lines}
Strong lines for determining the abundance of iron in {$\alpha$} Cen A were obtained from
Anstee, O'Mara \& Ross [9]. The A grade lines of iron from Anstee, O'Mara \& Ross
[9] have good f-­values and lead to an accurate abundance of iron in the Sun. Also the
Mg b-­lines, the Na D-­lines and the Ca I resonance line were used in the analysis.

\subsection{Atmospheric Model}
Four different model configurations based on the Holweger and M\"{u}ller model [25]
(hereafter referred to as the HM model) for the Sun were used in this analysis. The
{\bf first model} was the HM model with a solar composition and log {\it g} equal to 4.3 (as
calculated in section 6.2.2). In the {\bf second model} the abundances obtained from the
first were used to determine a new model which reflects the higher metallicity of {$\alpha$} Cen
A. To allow for the possibility that {$\alpha$} Cen may be bluer and hotter than the Sun the
temperature distribution in the second model was scaled to the effective temperature
of 5820 K ({\bf third model}) suggested by Bessell [10]. A {\bf fourth model} was then 
constructed from the third to incorporate the higher abundances resulting from the increase 
in the effective temperature.

\subsection{Abundance Analysis}
The spectrum synthesis program (SYN) was used for obtaining abundances of the
strong lines. This software enables the computed profile of the lines to be compared
with the observed line profile. In the software, the abundance was adjusted so as
to minimize the difference between the observed and the computed spectrum, which
is displayed on the computer screen scaled up by a factor of 5. The fitting process
concentrates on the damping wings of the line so that the derived abundance is
independent of the macroturbulent and microturbulent velocities.

The derived abundances obtained from the selected strong lines using the first
model are shown in Table 6.1. The difference in the last column is the difference
between {$\alpha$} Cen A and the solar abundances. The iron abundance obtained for {$\alpha$} Cen
A is compared with the iron abundance of the Sun obtained by Anstee, O'Mara \&
Ross [9]. The Mg, Na and Ca abundances for {$\alpha$} Cen A are compared with observed
solar abundances. For Ca, the difference of 0.14 dex is obtained by assuming a solar
abundance of Ca of 6.36 obtained using a selection of lines in the Sun. A solar
abundance based on only the $\lambda$4226.74 line comes out to be 6.38 and hence a lower
overabundance of 0.12 dex. Table 6.2 shows the mean abundances and the mean
overabundance compared to the solar abundances.

\renewcommand{\baselinestretch}{1}
\renewcommand{\arraystretch}{1.25}
\begin{table}[h]
\begin{center}
\parbox{14cm}{\caption {\it Abundance Analysis from Strong Lines of {$\alpha$} Cen A using {\bf the first
model}. The columns in order are: element, wavelength, log {\it gf}, line broadening
cross-­section $\sigma$, velocity exponent {$\alpha$} from the tables of Anstee \& O'Mara [5], reduced
equivalent width from Moore \etal [7], logarithmic abundance and the difference compared 
to observed solar abundances. Tables 6.3 and 6.5 have similar columns.}}
\begin{tabular}{|l|l|l|l|l|l|l|l|} \hline
Element &  $\lambda$ & log {\it gf} & $\sigma$ 
& $\alpha$ & $\Delta \lambda/ \lambda$ & logN/N(H) & Difference \\*[-0.15cm]
 & ({\AA}) &  & (a$^{2}_{0}$) &  &  & +12 & (dex) \\ \hline
Fe & 4071.738 & ­-0.022 & 332 & 0.250 & 191  & 7.60 & +0.10 \\
Fe & 4383.545 & 0.20  & 298 & 0.262 & 235   & 7.61 & +0.10 \\
Fe & 4415.123 & -­0.615 & 308 & 0.257 & 92.9 & 7.62 & +0.10 \\
Fe & 4918.994 ­& -0.340 & 750 & 0.237 & 53.7 & 7.64 & +0.13 \\
Fe & 4957.299 ­& -0.410 & 738 & 0.239 & 56.76 & 7.67 & +0.16 \\
Fe & 4957.597 & 0.230 & 724 & 0.240 & 128 & 7.67 & +0.16 \\
Fe & 5232.940 ­& -0.06 & 724 & 0.241 & 64.9 & 7.64 & +0.13 \\
Fe & 5269.537 ­& -1.321 & 239 & 0.249 & 87 & 7.63 & +0.12 \\
Fe & 5328.029 ­& -1.466 & 241 & 0.249 & 70.4 & 7.63 & +0.12 \\
Fe & 5328.532 ­& -1.850 & 240 & 0.257 & 39.4 & 7.63 & +0.11 \\
Fe & 5371.490 ­& -1.645 & 242 & 0.249 & 44.1 & 7.66 & +0.14 \\
Fe & 5446.917 ­& -1.86 & 243 & 0.249 & 42.8 & 7.62 & +0.11 \\
Mg & 5167.327 ­& -0.857 & 731 & 0.240 & 173 & 7.68 & +0.10 \\
Mg & 5172.698 ­& -0.380 & 731 & 0.240 & 234 & 7.68 & +0.10 \\
Mg & 5183.619 ­& -0.158 & 731 & 0.240 & 303 & 7.69 & +0.11 \\
Mg & 5528.905 ­& -0.470 & 1471 & 311 & 53.8 & 7.69 & +0.11 \\
Ca & 4226.740 & 0.243 & 372 & 242 & 342 & 6.50 & +0.12 \\
Na & 5889.973 & 0.1173 & 409 & 263 & 120 & 6.41 & +0.13 \\
Na & 5895.940 ­& -0.1838 & 409  & 237 & 91 & 6.43 & +0.15 \\ \hline
\end{tabular}
\end{center}
\end{table}
\renewcommand{\baselinestretch}{1.5}
\renewcommand{\baselinestretch}{1}
\renewcommand{\arraystretch}{1.25}
\begin{table}[h]
\begin{center}
\parbox{10cm}{\caption {\it Mean Abundances and Mean Overabundance for {$\alpha$} 
Cen A using {\bf the first model}.}}
\begin{tabular}{|l|l|l|} \hline
Element & Mean Logarithmic & Mean                    \\*[-0.15cm]
        & Abundance        & Overabundance (dex)     \\ \hline
Fe & 7.63 $\pm$ 0.02 & 0.12 $\pm$ 0.02 \\
Mg & 7.68 $\pm$ 0.01 & 0.10 $\pm$ 0.01 \\
Ca & 6.50 (1 line)   & 0.12 (1 line)   \\
Na & 6.42 $\pm$ 0.01 & 0.14 $\pm$ 0.01 \\ \hline
\end{tabular}
\end{center}
\end{table}
\renewcommand{\baselinestretch}{1.5}

The atmospheric model was revised to add the higher abundance obtained for
{$\alpha$} Cen A. The revised model was again used to get the abundances of the strong
lines. The abundances obtained using the second model are shown in Table 6.3.
Table 6.4 shows the mean abundance and mean overabundance for {$\alpha$} Cen A using
the higher abundance in the model (the second model). Here also the iron abundance
is compared with the solar iron abundance obtained by Anstee, O'Mara \& Ross [9].

\renewcommand{\baselinestretch}{1}
\renewcommand{\arraystretch}{1.25}
\begin{table}[h]
\begin{center}
\parbox{14cm}{\caption {\it Abundance Analysis from Strong Lines of {$\alpha$} 
Cen A using {\bf the second model}.}} 
\begin{tabular}{|l|l|l|l|l|l|l|l|} \hline
Element &  $\lambda$ & log {\it gf} & $\sigma$ 
& $\alpha$ & $\Delta \lambda/ \lambda$ & logN/N(H) & Difference \\*[-0.15cm]
 & ({\AA}) &  & (a$^{2}_{0}$) &  &  & +12 & (dex) \\ \hline
Fe & 4071.738 & ­-0.022 & 332 & 0.250 & 191 & 7.62 & +0.12 \\
Fe & 4383.545 & 0.20  & 298 & 0.262 & 235 & 7.63 & +0.12 \\
Fe & 4415.123 & -­0.615 & 308 & 0.257 & 92.9 & 7.64 & +0.12 \\
Fe & 4918.994 & -­0.340 & 750 & 0.237 & 53.7 & 7.67 & +0.16 \\
Fe & 4957.299 & -­0.410 & 738 & 0.239 & 56.76 & 7.71 & +0.20 \\
Fe & 4957.597 & 0.230 & 724 & 0.240 & 128 & 7.71 & +0.20 \\
Fe & 5232.940 & -­0.06 & 724 & 0.241 & 64.9 & 7.67 & +0.16 \\
Fe & 5269.537 & -­1.321 & 239 & 0.249 & 87 & 7.65 & +0.14 \\
Fe & 5328.029 & ­-1.466 & 241 & 0.249 & 70.4 & 7.65 & +0.14 \\
Fe & 5328.532 & ­-1.850 & 240 & 0.257 & 39.4 & 7.65 & +0.13 \\
Fe & 5371.490 & ­-1.645 & 242 & 0.249 & 44.1 & 7.69 & +0.17 \\
Fe & 5446.917 & ­-1.86 & 243 & 0.249 & 42.8 & 7.64 & +0.13 \\
Mg & 5167.327 & ­-0.857 & 731 & 0.240 & 173 & 7.74 & +0.16 \\
Mg & 5172.698 & ­-0.380 & 731 & 0.240 & 234 & 7.74 & +0.16 \\
Mg & 5183.619 & ­-0.158 & 731 & 0.240 & 303 & 7.74 & +0.16 \\
Mg & 5528.905 & ­-0.470 & 1471 & 311 & 53.8 & 7.74 & +0.16 \\
Ca & 4226.740 & 0.243 & 372 & 242 & 342 & 6.55 & +0.17 \\
Na & 5889.973 & 0.1173 & 409 & 263 & 120 & 6.48 & +0.20 \\
Na & 5895.940 & ­-0.1838 & 409 & 237 & 91 & 6.48 & +0.20 \\ \hline
\end{tabular}
\end{center}
\end{table}
\renewcommand{\baselinestretch}{1.5}
\renewcommand{\baselinestretch}{1}
\renewcommand{\arraystretch}{1.25}
\begin{table}[h]
\begin{center}
\parbox{10cm}{\caption {\it Mean Abundances and Mean Overabundance for {$\alpha$} Cen A using 
{\bf the second model}.}} 
\begin{tabular}{|l|l|l|} \hline
Element & Mean Logarithmic & Mean                     \\*[-0.15cm]
        & Abundance        & Overabundance (dex)      \\ \hline
Fe & 7.66 $\pm$ 0.03 & 0.15 $\pm$ 0.03 \\
Mg & 7.74 $\pm$ 0.00 & 0.16 $\pm$ 0.00 \\
Ca & 6.55 (1 line)   & 0.17 (1 line)   \\
Na & 6.48 $\pm$ 0.00 & 0.20 $\pm$ 0.00 \\ \hline
\end{tabular}
\end{center}
\end{table}
\renewcommand{\baselinestretch}{1.5}

The third revision of the model was done by scaling the temperatures by the ratio
of an estimated effective temperature of {$\alpha$} Cen A, 5820 K (from literature reviewed)
to the solar effective temperature, 5780 K (used in previous models). The basis of this
scaling follows from what one would expect for the case of a grey model atmosphere in
which the opacity is independent of wavelength, where, the temperature distribution
for {$\alpha$} Cen A is given by
\begin{equation}
T^4_{\alpha CenA}(\tau) = T^4_{e(\alpha CenA)} * (\tau + q(\tau))
\end{equation}
where, $\tau$ is the optical depth and q($\tau$) is a function which is chosen to make the flux 
independent of depth in the grey model atmosphere. The temperature distribution for the Sun is given by
\begin{equation}
T^4_\odot = T^4_{e\odot} * (\tau + q(\tau))
\end{equation}
Taking the ratio and the fourth root, we get,
\begin{equation}
T_{\alpha CenA} (\tau) = (T_{e(\alpha CenA)}/T_{e\odot}) * T_\odot (\tau)
\end{equation}

Abundances obtained using the model with the higher effective temperature after
iterating the composition to reflect the higher temperature, i.e. model 4, are shown
in Table 6.5 and their average values are shown in Table 6.6.

\renewcommand{\baselinestretch}{1}
\renewcommand{\arraystretch}{1.25}
\begin{table}[h]
\begin{center}
\parbox{14cm}{\caption {\it Abundance Analysis from Strong Lines of {$\alpha$} Cen A 
using {\bf the third and the fourth model}.}} 
\begin{tabular}{|l|l|l|l|l|l|l|l|} \hline
Element &  $\lambda$ & log {\it gf} & $\sigma$ 
& $\alpha$ & $\Delta \lambda/ \lambda$ & logN/N(H) & Difference \\*[-0.15cm] 
 & ({\AA}) &  & (a$^{2}_{0}$) &  &  & +12 & (dex) \\ \hline
Fe & 4071.738 & -­0.022 & 332 & 0.250 & 191 & 7.67 & +0.17 \\
Fe & 4383.545 & 0.20 & 298 & 0.262 & 235 & 7.70 & +0.19 \\
Fe & 4415.123 & -­0.615 & 308 & 0.257 & 92.9 & 7.70 & +0.18 \\
Fe & 4918.994 & -­0.340 & 750 & 0.237 & 53.7 & 7.74 & +0.22 \\
Fe & 4957.299 & ­-0.410 & 738 & 0.239 & 56.76 & 7.77 & +0.26 \\
Fe & 4957.597 & 0.230 & 724 & 0.240 & 128 & 7.77 & +0.26 \\
Fe & 5232.940 & ­-0.06 & 724 & 0.241 & 64.9 & 7.70 & +0.19 \\
Fe & 5269.537 & ­-1.321 & 239 & 0.249 & 87 & 7.69 & +0.18 \\
Fe & 5328.029 & ­-1.466 & 241 & 0.249 & 70.4 & 7.70 & +0.19 \\
Fe & 5328.532 & ­-1.850 & 240 & 0.257 & 39.4 & 7.71 & +0.19 \\
Fe & 5371.490 & ­-1.645 & 242 & 0.249 & 44.1 & 7.72 & +0.20 \\
Fe & 5446.917 & ­-1.86 & 243 & 0.249 & 42.8 & 7.69 & +0.18 \\
Mg & 5167.327 & ­-0.857 & 731 & 0.240 & 173 & 7.78 & +0.20 \\
Mg & 5172.698 & ­-0.380 & 731 & 0.240 & 234 & 7.79 & +0.21 \\
Mg & 5183.619 & ­-0.158 & 731 & 0.240 & 303 & 7.80 & +0.22 \\
Mg & 5528.905 & ­-0.470 & 1471 & 311 & 53.8 & 7.78 & +0.20 \\
Ca & 4226.740 & 0.243 & 372 & 242 & 342 & 6.59 & +0.21 \\
Na & 5889.973 & 0.1173 & 409 & 263 & 120 & 6.53 & +0.25 \\
Na & 5895.940 & ­-0.1838 & 409 & 237 & 91 & 6.53 & +0.25 \\ \hline
\end{tabular}
\end{center}
\end{table}
\renewcommand{\baselinestretch}{1.5}

\renewcommand{\baselinestretch}{1}
\renewcommand{\arraystretch}{1.25}
\begin{table}[h]
\begin{center}
\parbox{10cm}{\caption {\it Mean Abundances and Mean Overabundance for {$\alpha$} Cen A using 
{\bf the third and the fourth model}.}} 
\begin{tabular}{|l|l|l|} \hline
Element & Mean Logarithmic & Mean                \\*[-0.15cm]
        & Abundance        & Overabundance (dex) \\ \hline
Fe & 7.71 $\pm$ 0.03 & 0.20 $\pm$ 0.03 \\
Mg & 7.79 $\pm$ 0.01 & 0.21 $\pm$ 0.01 \\
Ca & 6.59 (1 line)   & 0.21 (1 line)   \\
Na & 6.53 $\pm$ 0.00 & 0.25 $\pm$ 0.00 \\ \hline
\end{tabular}
\end{center}
\end{table}
\renewcommand{\baselinestretch}{1.5}

Figures 6.1, 6.2 and 6.3 are self explanatory. The observed line profile is shown by
a solid line and the computed line profile is represented by dots. Figure 6.1 shows the
$\lambda$4226.74 resonance line of neutral calcium. This very strong line, synthesised with
numerous blending lines, give a quite accurate calcium abundance. As only this line
is used, the abundance comparison and hence the overabundance is calculated with
respect to this line only. The second plot shows the $\lambda$5895.94 sodium ($D_2$) line. Due
to the fact that Mt. Stromlo is neither high nor dry the wings suffer from considerable
contamination by telluric water vapour lines causing uncertainty in determination of
the overabundance.

Figure 6.2 shows how the process of determining abundance by strong lines is
important using spectrum synthesis. The first plot gives the best fit to the $\lambda$5269.53
iron line. The neighbouring line is a medium-­strong calcium line. The second plot
shows the variation in the wings of this line when abundance is changed by $\pm$0.1 dex.
Matching the wings of this line results in an abundance independent of microturbulence.

Figure 6.3 shows the $\lambda$5172.69 magnesium b-­line. It also shows the neighbouring
medium-­strong lines of iron. The second plot shows $\lambda$5328.02 and $\lambda$5328.53 pair of
iron lines.

\begin{figure}[ht]
\hspace{2.5cm}
\psfig{file=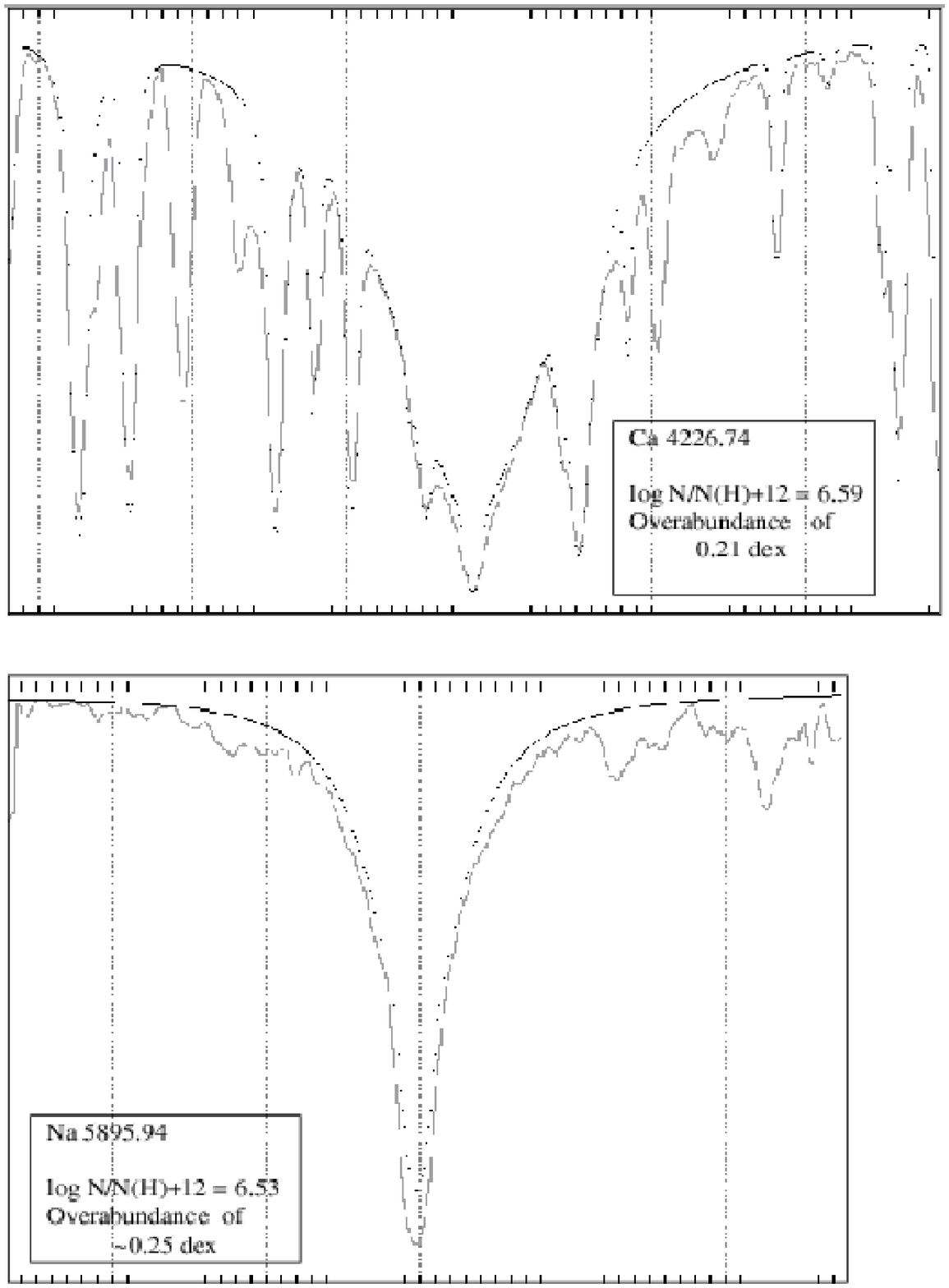,height=18cm,width=10cm}
\caption{\it 
Resonance line of neutral calcium $\lambda$4226.74 (top) and Sodium ($D_2$) line
$\lambda$5895.94 (bottom). The vertical scale is the residual intensity with a residual intensity
of zero at the bottom of each plot and unity just below the top of the plots. The
horizontal scale is the wavelength, with the angstr\"{o}m markers extending from the top
to the bottom of the plots with one-­tenth angstr\"{o}m markers in between. Figures 6.2
and 6.3 have similar vertical and horizontal scales.}
\end{figure}

\begin{figure}[ht]
\hspace{2.5cm}
\psfig{file=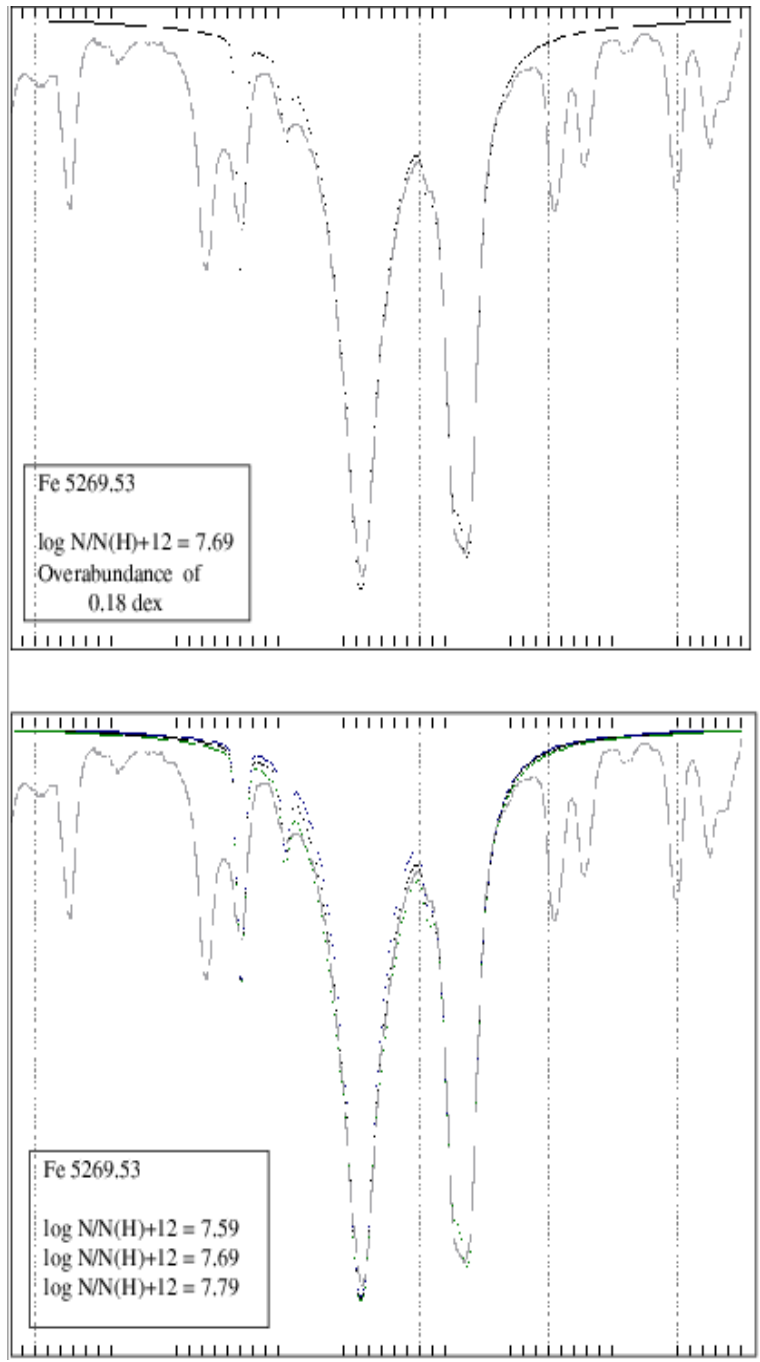,height=18cm,width=10cm}
\caption{\it 
The upper plot shows the observed profile (line) for the $\lambda$5269.53 line
and the computed profile (dots) that best fits the observed line profile. The lower plot
shows the fit for an abundance 0.1 dex lower and 0.1 dex higher than the abundance
that best fits the line profile. This lower plot clearly identifies the damping wings of
the line and the core, which is dominated by photospheric motions.}
\end{figure}

\begin{figure}[ht]
\hspace{2.5cm}
\psfig{file=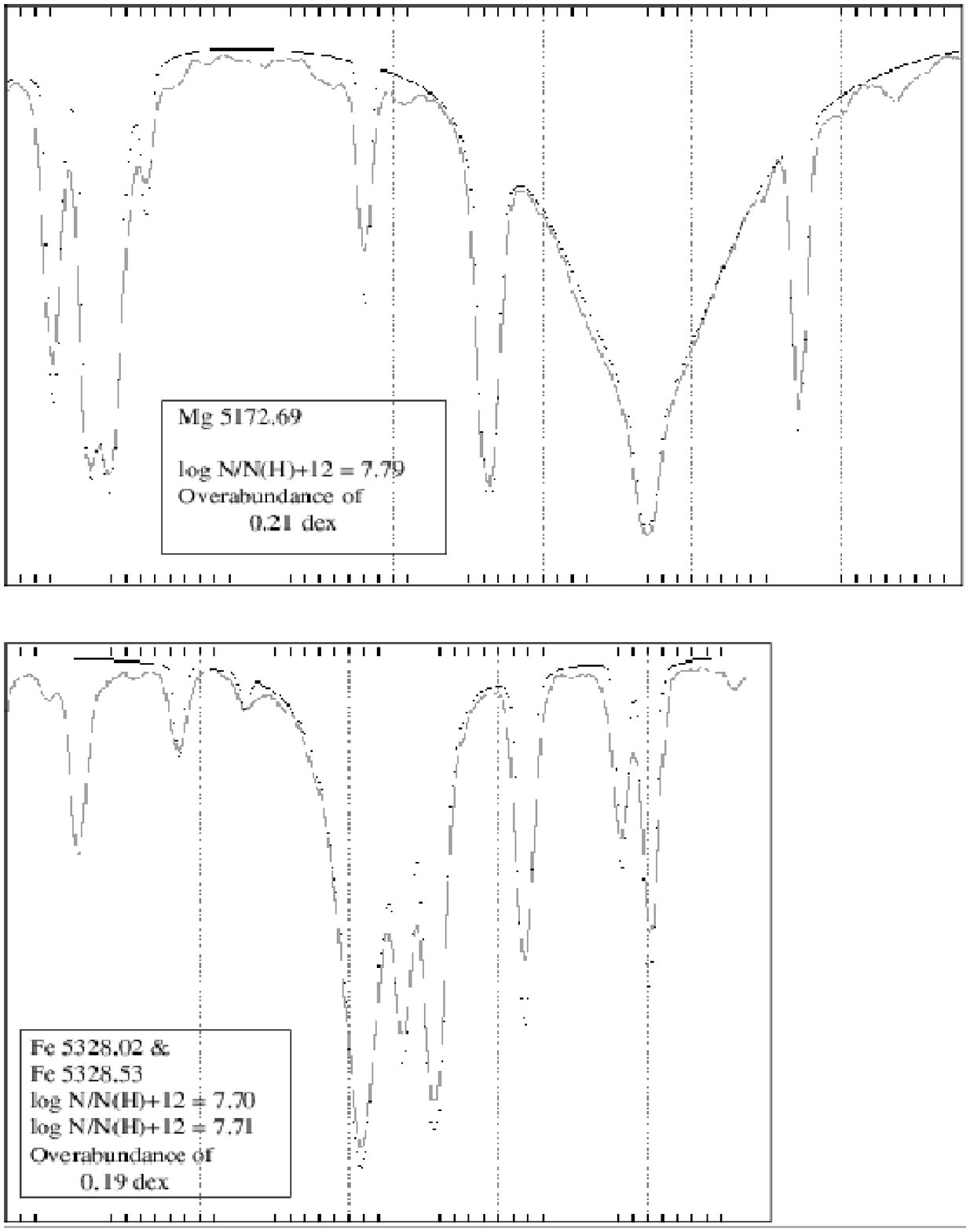,height=18cm,width=10cm}
\caption{\it Magnesium b-­line $\lambda$5172.69 (top) and Iron lines $\lambda$5328.02 and  $\lambda$5328.53
(bottom)}
\end{figure}

{\bf The results obtained for $\alpha$ Cen A in Table 6.6 are the final results for its chemical
analysis. The results clearly show that $\alpha$ Cen A is a metal rich star compared to the
Sun}. A few important points to be noted from this analysis are as follows:
\begin{itemize}
\item The abundance of metals other than iron are obtained using the wings of very
few lines. The Mg abundance obtained from four strong lines of magnesium is
quite accurate, the Ca abundance could be obtained from the only strong line
available, and the Na abundance obtained from two strong lines is likely to be
overestimated because of presence of water vapour lines in the wings of these
two lines as shown in Figure 6.1.
\item The effect of temperature on iron lines was noticeable, raising the temperature
raises the computed abundance, with some lines more affected than others. The
increase in the derived abundance was particularly noticeable for the iron line
pair of $\lambda$4957.29 and $\lambda$4957.59.
\item The value of log {\it g} is known with an accuracy of at least 10\% and therefore should
not, as was the case in the analysis of Furenlid \& Meylan [4], be regarded as an
adjustable parameter in the analysis. The only problem in the model seems to
be the value of the effective temperature. When medium-­strong lines are used
in an extension of this analysis one could use variations in the abundance of
an element with excitation potential to aid in the determination of the most
suitable effective temperature.
\end{itemize}

\section{The Determination of the Microturbulence from Medium--Strong Lines of Iron in {$\alpha$} Cen A}
\subsection{Selection of Lines}
Medium--strong lines having equivalent widths in the range of about 60--100 m{\AA} were
used for estimating the microturbulence in {$\alpha$} Cen A. The lines of Blackwell \etal [22],
which were used in Chapter 4 for solar microturbulence, are used here for {$\alpha$} Cen A.
The equivalent widths for most of the lines were available from Furenlid \& Meylan
[4], while some of the lines were used from our own data, approximating the equivalent
widths by fitting computed and observed lines in the spectrum synthesis program. Few lines 
which were not in the Furenlid \& Meylan [4] data and in our own data, were
omitted in this analysis.

\subsection{Microturbulence from Medium--Strong Lines of Iron}
The first analysis was done using the model of Holweger­M\"{u}ller [25] of the Sun with
log {\it g} = 4.3 and an abundance higher by 0.15 dex (the second model). Keeping
the abundance fixed at 7.66 for all medium-­strong lines, the equivalent width was
adjusted to match the observed equivalent width by varying the microturbulence.
The process was done using the SYN program of John Ross [2]. Table 6.7 shows the
microturbulence obtained using this process. The line broadening cross­sections ($\sigma$)
and velocity parameters ($\alpha$) were obtained by interpolation in the tables of Anstee \&
O'Mara [5]. The mean value of microturbulence obtained from this analysis is 1.35
$\pm$ 0.13 km/s.

The process was repeated using the atmospheric model with higher effective temperature 
and incorporated higher abundance (the fourth model). Table 6.8 shows the
microturbulence obtained using higher effective temperature and hence the higher 
abundance of 7.71 (0.20 dex higher). The effect of the temperature is clearly visible
from this part of the analysis. As the temperature increases, the abundance from
strong lines increases and consequently the microturbulence from medium--strong lines
decreases. The mean microturbulence obtained with higher temperature and higher
abundance is 1.34 $\pm$ 0.13 km/s.
\renewcommand{\baselinestretch}{1}
\renewcommand{\arraystretch}{1.25}
\begin{table}[ht]
\begin{center}
\parbox{15cm}{\caption {\it Microturbulence Analysis of Fe I Lines in {$\alpha$} Cen A 
using {\bf the second model}. The line broadening cross-­sections $\sigma$ and velocity 
parameters {$\alpha$} were obtained
by interpolation in the tables of Anstee and O'Mara [5]. The microturbulence required
to match the observed equivalent widths were obtained using the spectrum synthesis
program of Ross [2]. '*' are equivalent widths estimated from the available data. The
values are crudely approximated by fitting the computed spectrum to the observed
spectrum. Other equivalent widths are taken from Furenlid \& Meylan [4]. Table 6.8
has similar columns.}} 
\begin{tabular}{|l|l|l|l|l|l|} \hline
Wavelength & Equivalent & Broadening & Enhance- & Velocity & Microturbulence \\*[-0.15cm]
({\AA}) & Width & Cross- & ment & Parameter & ($\xi$ km/s) {\bf with} \\*[-0.15cm]
        & (m{\AA}) & section $\sigma$ & Factor & {$\alpha$} & {\bf an abundance} \\*[-0.15cm]
        &          & (a$^{2}_{0}$) &  &  & {\bf 7.66} \\ \hline
4389.24 & 82.00* & 218 & 2.13 & 0.250 & 1.212 \\
5247.06 & 74.00* & 206 & 3.30 & 0.254 & 1.266 \\
5250.21 & 73.00* & 208 & 3.22 & 0.254 & 1.311 \\
5701.55 & 96.00* & 364 & 2.05 & 0.241 & 1.325 \\
5956.70 & 56.00  & 229 & 2.54 & 0.252 & 1.205 \\
6151.62 & 59.00  & 280 & 1.62 & 0.264 & 1.385 \\
6173.34 & 75.00  & 283 & 1.62 & 0.267 & 1.175 \\
6200.32 & 83.00  & 354 & 2.20 & 0.240 & 1.379 \\
6265.14 & 101.00 & 277 & 1.63 & 0.262 & 1.479 \\
6297.80 & 86.00  & 280 & 1.63 & 0.264 & 1.322 \\
6593.88 & 101.00 & 324 & 2.43 & 0.248 & 1.596 \\
6609.12 & 77.00  & 337 & 2.37 & 0.246 & 1.572 \\
6750.15 & 85.00  & 282 & NA   & 0.260 & 1.361 \\ \hline
\end{tabular}
\end{center}
\end{table}
\renewcommand{\baselinestretch}{1.5}

\renewcommand{\baselinestretch}{1}
\renewcommand{\arraystretch}{1.25}
\begin{table}[h]
\begin{center}
\parbox{15cm}{\caption {\it Microturbulence Analysis of Fe I Lines in {$\alpha$} Cen A 
using {\bf the fourth model}.}} 
\begin{tabular}{|l|l|l|l|l|l|} \hline
Wavelength & Equivalent & Broadening & Enhance- & Velocity & Microturbulence \\*[-0.15cm]
({\AA}) & Width & Cross- & ment & Parameter & ($\xi$ km/s) {\bf with} \\*[-0.15cm]
        & (m{\AA}) & section $\sigma$ & Factor & {$\alpha$} & {\bf an abundance} \\*[-0.15cm]
        &          & (a$^{2}_{0}$) &  &  & {\bf 7.71} \\ \hline
4389.24 & 82.00* & 218 & 2.13 & 0.250 & 1.225 \\
5247.06 & 74.00* & 206 & 3.30 & 0.254 & 1.278 \\
5250.21 & 73.00* & 208 & 3.22 & 0.254 & 1.320 \\
5701.55 & 96.00* & 364 & 2.05 & 0.241 & 1.315 \\
5956.70 & 56.00  & 229 & 2.54 & 0.252 & 1.199 \\
6151.62 & 59.00  & 280 & 1.62 & 0.264 & 1.346 \\
6173.34 & 75.00  & 283 & 1.62 & 0.267 & 1.160 \\
6200.32 & 83.00  & 354 & 2.20 & 0.240 & 1.360 \\
6265.14 & 101.00 & 277 & 1.63 & 0.262 & 1.474 \\
6297.80 & 86.00  & 280 & 1.63 & 0.264 & 1.312 \\
6593.88 & 101.00 & 324 & 2.43 & 0.248 & 1.586 \\
6609.12 & 77.00  & 337 & 2.37 & 0.246 & 1.542 \\
6750.15 & 85.00  & 282 & NA   & 0.260 & 1.346 \\ \hline
\end{tabular}
\end{center}
\end{table}
\renewcommand{\baselinestretch}{1.5}

\chapter{Conclusions}
The abundances of iron, magnesium, calcium and sodium in {$\alpha$} Cen A have been
determined from strong lines and are approximately 0.2 dex higher than in the Sun. As these
derived abundances are independent of non-­thermal motions in the atmosphere of $\alpha$
Cen A, the metal rich status of this star is unambiguous since the higher abundances
derived cannot be reduced to solar values by adopting a higher microturbulence.

The microturbulence obtained from medium-­strong lines using the abundance
obtained from the strong lines is 1.34 $\pm$ 0.12 km/s, i.e., 0.20 km/s larger than in the
Sun, which is fully consistent with the lower surface gravity of {$\alpha$} Cen A. This analysis
gives a strong hold on the microturbulence making it possible to derive improved
abundances of other elements for which there are no strong lines.

Future work should include:
\begin{itemize}
\item extension of the abundance determination to other elements,
\item acquisition of the spectrum in the region from 6000 to 7000 {\AA}, which will provide
access to additional strong lines of calcium and other elements,
\item better determination of the angular diameter from measurements made by the
Sydney University Stellar Interferometer (SUSI) leading to a better determination of 
the surface gravity, and
\item also spectro-­photometry would lead to a better determination of the effective
temperature.
\end{itemize}
Combined with even higher \SN data, it should be possible to obtain excellent abundances 
for {$\alpha$} Cen A. Assuming that the metal abundance for {$\alpha$} Cen B is the same as
for {$\alpha$} Cen A, it should be possible to learn a great deal about stars somewhat cooler
than the Sun from a similar analysis of {$\alpha$} Cen B.

In summary, for the first time, an unambiguous microturbulence and metallicity
for {$\alpha$} Cen A has been obtained !

\chapter*{Appendix}
\addcontentsline{toc}{chapter}{Appendix}

\section*{{$\alpha$} Cen -­ a Candidate for Terrestrial Planets and
Intelligent Life}
While this is outside the general thrust of the thesis, it may be of interest to note
other reasons why the {$\alpha$} Cen system has received so much attention.

\section*{Why {$\alpha$} Cen is Special?}
The nature of the star around which a planet orbits is an important factor in 
determining whether life is possible on the planet. Seventy percent of all stars in the
Galaxy are red dwarfs like Proxima Cen, too faint, too cool and in some cases too
variable to support life. About 15 percent are orange K-­type dwarfs. Although the
more luminous K dwarfs like {$\alpha$} Cen B, may be bright and warm enough for life but
the fainter ones in the class may be too dim and cool. Another 10 percent of stars
are white dwarfs -­ dying stars that either could not have life, or must have destroyed
any life they once had. That leaves the brightest 5 percent of all stars in the Galaxy,
a privileged group to which both the Sun and {$\alpha$} Cen A belong. Most of this upper 5
percent consists of yellow G-­type stars that are bright, warm, and good for life.

Hence it is very important to study these type of stars as they might have intelligent 
life around them. 

\section*{Tests for Possibility of Life}
A star must pass five tests before it becomes a promising place for existence of life
[8]. They are :
\begin{itemize}
\item The star should be on the main sequence to ensure maturity and stability.
\item Spectral type of the star.
\item A system must demonstrate stable conditions.
\item The star's age -­ a star must be old enough to give life a chance to evolve.
\item Chemical composition of the star -­ does the star have the heavy elements that
biological life needs ?
\end{itemize}
The first criterion is to ensure the star's maturity and stability, which means it has
to be on the main sequence. Main sequence stars fuse hydrogen into helium in their
cores, generating light and heat. Because hydrogen is so abundant in stars, most of
them stay on the main sequence a long time, giving life a chance to evolve. The Sun
and all three components of {$\alpha$} Cen pass this test.

The second test is much tougher. The star must have the right spectral type
because this determines how much energy a star emits. The hotter stars those with
spectral types O, B, A, and early F are no good because they burn out fast and die
quickly, emitting ultra--violet rays. The cooler stars ­ those with spectral types M
and late K ­may not produce enough energy to sustain life, for instance they may not
permit the existence of liquid water on their planets. Between the stars that are too
hot and those that are too cool, yellow G type stars like the Sun can give rise to life.
{$\alpha$} Cen A passes this test as it is of the same class as our Sun. {$\alpha$} Cen B is
a K1 star, so it is hotter and brighter than most K stars, therefore it may pass this
test or it may not. The red dwarf Proxima fails this test.

For the third test, a system must demonstrate stable conditions, as {$\alpha$} Cen A and
B form a binary pair how much does the light received by the planets of one star
vary as the other star revolves around it? During their 80-­year orbit, the separation
between A and B changes from 11 AU to 35 AU. As viewed from the planets of one
star, the brightness of the other increases as the stars approach and decrease as the
stars recede. Fortunately, the variation is too small to matter, and {$\alpha$} Cen A and B
pass this test. Proxima being a red dwarf, which emits flares that causes its light to
double or triple in just few minutes, fails this test.

The fourth condition concerns the star's age. The Sun is about 4.6 billion years
old, so life on earth had enough time to evolve. A star must be old enough to give
life a chance to evolve. Remarkably, {$\alpha$} Cen A \& B are probably even older than the
Sun, they have an age of about 5 to 6 billion years, therefore they pass this test, but
the study of Proxima shows it to be younger compared to {$\alpha$} Cen A and B, so it fails
this test, too.

And the fifth and final test: Do the stars have enough heavy elements such as
carbon, nitrogen, oxygen and iron, to support biological life? Like most stars, the
Sun is primarily hydrogen and helium, but 2 percent of the Sun's mass is metals.
Although 2 percent may not sound a lot, it is enough to build rocky planets and to
give rise to us. And again, fortunately, {$\alpha$} Cen A and B pass this test. They are
metal-­rich stars.

Thus we see that {$\alpha$} Cen A passes all five tests, {$\alpha$} Cen B passes all but one, and
only Proxima Cen could not satisfy most. The details are summarised in the given Table A.1.

\renewcommand{\baselinestretch}{1}
\renewcommand{\arraystretch}{1.25}
\begin{table}[ht]
\begin{center}
{Table A.1: \it Terrestrial Life Conditions ­Important Questions [6]}. \\
\begin{tabular}{|l|l|l|l|l|} \hline
Terrestrial Life Conditions: & Sun & {$\alpha$} Cen A & {$\alpha$} Cen B & Proxima \\*[-0.15cm]
Questions for Any Star & & & & \\ \hline
On the main sequence? & Yes & Yes & Yes & Yes \\
Of the right spectral type? & Yes & Yes & Maybe & No \\
Constant in brightness? & Yes & Yes & Yes & No \\
Old enough? & Yes & Yes & Yes & No? \\
Rich in metals? & Yes & Yes & Yes & ? \\
Has stable planetary orbits? & Yes & Yes & Yes & Yes \\
Could planets form? & Yes & ? & ? & ? \\
Do planets actually exist? & Yes & ? & ? & ? \\
Small rocky planets possible? & Yes & Yes & Yes & Yes? \\
Planets in the life zone? & Yes & Maybe & Maybe & No \\ \hline
\end{tabular}
\end{center}
\end{table}
\renewcommand{\baselinestretch}{1.5}

As stars of {$\alpha$} Cen, specially {$\alpha$} Cen A and B, are very similar to the Sun and
probably have the same age as the Sun, the most likely question could be, Is there
any possibility of an earth-­like planet revolving around {$\alpha$} Cen in stable orbit? The
possible interpretation given by Soderblom [27] in favour of {$\alpha$} Cen A goes something
like this:
\begin{quote}
Three body systems are unstable unless two of the bodies are close together and
the third is a large distance away. If the separation of the three are comparable, it
doesn't take long for two of them to pass close to each other, which sends at least
one of them streaking off into the void, no longer bound to the other two bodies.
\end{quote}
The distance between {$\alpha$} Cen A and B varies from about 11 to 35 AU over the
course of its 80 year orbit. An earth--like planet 1 AU from {$\alpha$} Cen A could probably
stay there indefinitely without fear of harm from {$\alpha$} Cen B. According to Soderblom
[27] the only question is whether an earth-­like planet could form in the first place
at that location. Probabilistic discussion of whether planets like earth revolve round
the {$\alpha$} Cen A or B is given by Croswell [6]. At the end of his discussion, Croswell [6]
\begin{figure}[h]
\hspace{1cm}
\psfig{file=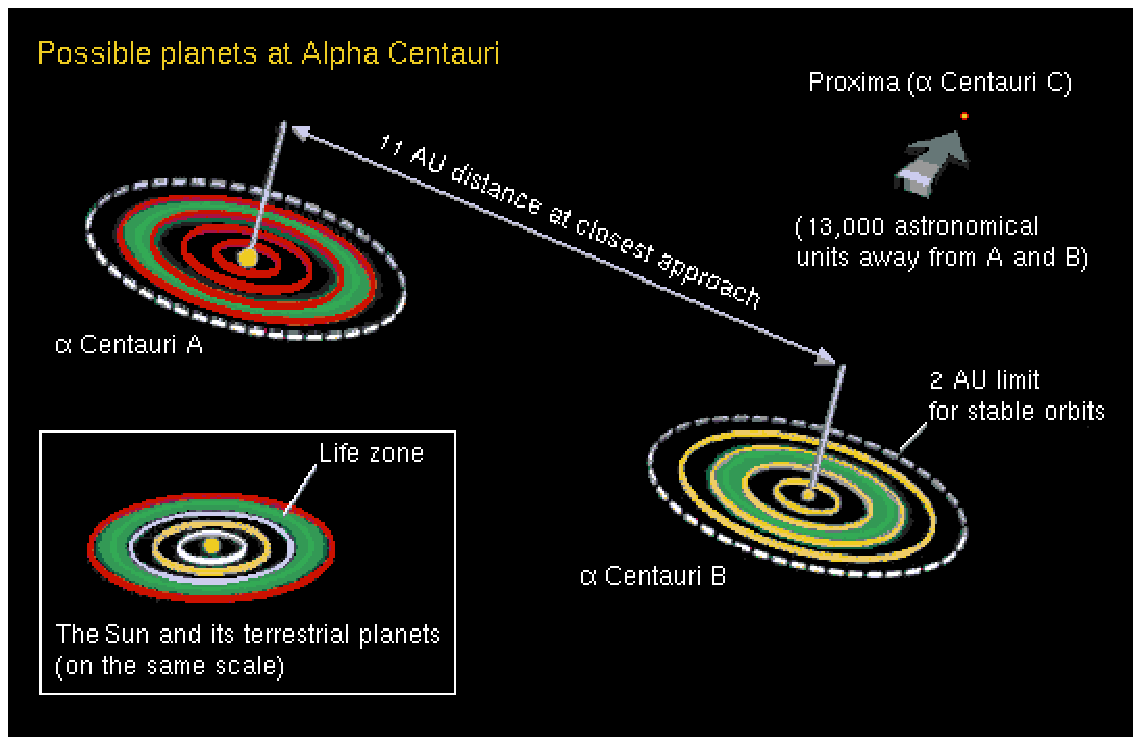,height=10.0cm,width=13.0cm}
\begin{center}Figure A.1: {\it Possibility of Planet in Life Zone of {$\alpha$} Cen [8].}
\end{center}
\end{figure}
concludes that it is frustrating that we can't reach any definite conclusion. Though
we can say with certainty that at least one of its star is ideal, we can say very little
about any planets there. On the positive side, earth-­like planets can have stable,
earth-­like orbits around either A or B. What is lacking is any direct evidence that
such planets exist.

{$\alpha$} Centauri is therefore very special; special not only for its proximity but
also for its promise. If {$\alpha$} Cen has planets, it is even more special since one or more
of those planets could resemble Earth. So if we ever launch missions to other stars in
the Galaxy, {$\alpha$} Cen will certainly be our first target.

\setlength{\parindent}{0cm}

\chapter*{Bibliography}
\addcontentsline{toc}{chapter}{Bibliography}

[1] S. D. Anstee and B. J. O'Mara. An investigation of Brueckner's theory of line
broadening with application to the sodium D lines. {\it Monthly Notices of the Royal
Astronomical Society}, 253:549--560, July 1991.

[2] J. E. Ross. Homepage for spectrum synthesis program (SYN). \\ 
Web Site -- {\it http://www.physics.uq.edu.au/people/ross/syn.htm}

[3] Y. Chmielewski, E. Friel, G. Cayrel de Strobel, and C. Bentolila. The 1992 
detailed analyses of {$\alpha$} Centauri A and {$\alpha$} Centauri B. 
{\it Astronomy and Astrophysics}, 263:219--231, May 1992.

[4] I. Furenlid and T. Meylan. An abundance analysis of {$\alpha$} Centauri A. 
{\it The Astrophysical Journal}, 350:827--838, February 1990.

[5] S. D. Anstee and B. J. O'Mara. Width cross­sections for collisional broadening
of s-­p and p-­s transitions by atomic hydrogen. {\it Monthly Notices of the Royal
Astronomical Society}, 276:859--866, April 1995. 

[6] K. Croswell. Does alpha Centauri have intelligent life? {\it Astronomy}, 19(4):28--37,
1991.

[7] C. E. Moore, M. G. J. Minnaert, and J. Houtgast. The Solar Spectrum 2935 {\AA}
to 8770 {\AA} : {\it Second Revision of Rowland's Preliminary Table of Solar Spectrum
Wavelengths}. Washington : National Bureau of Standards, 1966.

[8] Alpha Centauri. Web Site -­ {\it http://monet.physik.unibas.ch/¸schatzer/Alpha­
Centauri.html}, 1996.

[9] S. D. Anstee, B. J. O'Mara, and J. E. Ross. A determination of the solar
abundance of iron from the strong lines of Fe I. {\it Monthly Notices of the Royal
Astronomical Society}, 284:202--212, January 1997.

[10] M. S. Bessell. Alpha Centauri. {\it Proceedings Astronomical Society of Australia},
4(2):212--214, 1981.

[11] V. A. French and A.L.T. Powell. Title unknown. {\it Royal Observatory Bulletin},
173:63, 1971.

[12] D. E. Blackwell and M. J. Shallis. Stellar angular diameters from infrared 
photometry. Application to Arcturus and other stars; with effective temperatures.
{\it Monthly Notices of the Royal Astronomical Society}, 180:177--191, February 1977.

[13] D. R. Soderblom and D. Dravins. High resolution spectroscopy of Alpha Centauri. 
I. Lithium depletion near one solar mass. {\it Astronomy and Astrophysics},
140:427--430, July 1984.

[14] D. R. Soderblom. The temperatures of Alpha Centauri A and B. {\it Astronomy
and Astrophysics}, 158:273--274, 1986.

[15] A. Noels, N. Grevesse, P. Magain, C. Neuforge, A. Baglin, and Y. Lebreton.
Calibration of the {$\alpha$} Centauri system: metallicity and age. {\it Astronomy and
Astrophysics}, 247:91--94, January 1991.

[16] C. Neuforge. Alpha Centauri revisited. {\it Astronomy and Astrophysics}, 
268:650--652, 1993.

[17] J. Fernandes and C. Neuforge. {$\alpha$} Centauri and convection theories. {\it Astronomy
and Astrophysics}, 295:678--684, 1995.

[18] S. D. Anstee, B. J. O'Mara, and J. E. Ross, editors. The Broadening of Metallic
Lines in Cool Stars, January 1997.

[19] S. D. Anstee. The Collisional Broadening of Alkali Spectral Lines by Atomic
Hydrogen. {\it PhD thesis}, University of Queensland, Department of Physics, 1992.

[20] P. Barklem and B. J. O'Mara. The broadening of p-­d and d-­p transitions by collision 
with neutral hydrogen atoms. {\it Monthly Notices of the Royal Astronomical
Society}, 1997. Submitted.

[21] H. Holweger. Solar element abundances, Non-­LTE line formation in Cool stars
and Atomic data. {\it Physica Scripta}, T65:151--157, 1996.

[22] D. E. Blackwell, A. E. Lynas-­Gray, and G. Smith. On the determination of the
solar iron abundance using Fe I lines. {\it Astronomy and Astrophysics}, 296:217--232,
1995.

[23] R. L. Kurucz, I. Furenlid, J. Brault, and L. Testerman. {\it The solar Flux Atlas
from 296 to 1300 nm}. National Solar Observatory, Sunspot, New Mexico, 1984.

[24] J. E. Ross and L. H. Aller. The chemical composition of the Sun. {\it Science},
191:1223--1229, 1976.

[25] H. Holweger and E. A. M\"{u}ller. The photospheric barium spectrum: Solar abundance 
and collision broadening of Ba II lines by hydrogen. {\it Solar Physics}, 
39:19--30, 1974.

[26] C. Turon, D. Morin, F. Arenou, and M. A. C. Perryman. {\it CD-­ROM version of
the Hipparcos Input Catalogue HICIS software}. CD-­ROM, 1997.

[27] D. R. Soderblom. The Alpha Centauri system. {\it Mercury}, pages 138--140,
September-­October 1987.

\end{document}